\documentclass[12pt]{article}
\usepackage{amssymb}
\usepackage{amsmath}
\usepackage{graphicx}
\newcommand{\be}{\begin{equation}}
\newcommand{\ee}{\end{equation}}
\newcommand{\bea}{\begin{eqnarray}}
\newcommand{\eea}{\end{eqnarray}}
\newcommand{\dn}{\mathop{\mathrm{dn}}\nolimits}

\begin{document}

\title{\textbf{Non-minimally Coupled Cosmological Models with the Higgs-like Potentials and Negative Cosmological Constant}}

\author{I.Ya.~Aref'eva$^1$\footnote{arefeva@mi.ras.ru}, \ N.V.~Bulatov$^2$\footnote{nick\_bulatov@mail.ru}, \
R.V.~Gorbachev$^1$\footnote{rgorbachev@mi.ras.ru}, \ S.Yu.~Vernov$^3$\footnote{svernov@theory.sinp.msu.ru}\\[2.7mm]
\small{${}^1$Steklov Mathematical Institute, Russian Academy of Sciences,}\\
\small{Gubkina st. 8, 119991, Moscow, Russia}\\
\small{${}^2$Faculty of Physics, Lomonosov Moscow State University,}\\
\small{Leninskie Gory 1, 119991, Moscow, Russia,}\\
\small{${}^3$Skobeltsyn Institute of Nuclear Physics, Lomonosov  Moscow State University,}\\
\small{Leninskie Gory 1, 119991, Moscow, Russia}
}
\date { \ }
\maketitle

\begin{abstract}
We study dynamics of non-minimally coupled scalar field cosmological models  with Higgs-like potentials and a negative cosmological constant.
In these models the inflationary stage of the Universe evolution changes into a quasi-cyclic stage of the Universe evolution with oscillation behaviour of the Hubble parameter from positive to negative values. Depending on the initial conditions the Hubble parameter can perform either one or
several cycles  before to become negative forever.
\end{abstract}

\maketitle

\section{Introduction}
Models with scalar fields play a central role in the current description of the evolution of the Universe at the early epoch \cite{Starobinsky:1979ty,Mukhanov:1981xt,Guth:1980zm,Linde:1981mu,Albrecht:1982wi,Linde:1983gd}.
However,  predictions of  simplest inflationary models are  in
disagreement with the Planck2013 results~\cite{PlanckInflation,Planck2013}. At the same time many of these inflationary scenarios can be improved by adding a tiny non-minimal coupling of the inflaton field to gravity~\cite{GB2013,KL2013}.
The models with the Ricci scalar multiplied by a function of the scalar field are being intensively studied in the inflationary cosmology~\cite{SBB,nonmin-inf,Kaiser:1994,Kaiser:1994vs,HI,Bezrukov:2008ut,induced,HI1,HI3,HI4,Bezrukov2013}. In the last years,  the Higgs-driven inflation has attracted a lot of attention~\cite{HI,Bezrukov:2008ut,HI1,HI3,HI4,Bezrukov2013}.
The models, which include both the Higgs boson and dilaton~\cite{Higgs-Dilaton}, have been proposed  to describe not only an inflationary stage, but also a late period of Dark Energy domination.
All these models deal with positive definite potentials.

The goal of this paper is to explore new mathematical features of cosmological
models with non-minimally coupled scalar fields in the case of potentials with
negative minima.
At the present moment we cannot say that there is any necessity to consider such
 types of cosmological models. However, this type of models can appear as a  result
of renormalization, so, our consideration can give restrictions to a choice of the
realistic models.
Also as it has been shown~\cite{ABG}, non-positive definite potentials can arise as a consequence of nonlocality.
Theories with nonlocal scalar fields \cite{Aref'eva:2004qr}, inspired by the
string field theory \cite{Arefeva:2001ps} have been considered  as early Universe
models, either inflationary models~\cite{Lidsey,BarCline,Aref'eva:2011gj} or models
with bounce solutions~\cite{Aref'eva:2005gg,Aref'eva:2007uk}. Perturbations in such
models have been studied in~\cite{KV,Galli:2010qx}. In one of these models the scalar
field is the tachyon of the NSR fermion string and the model has the form of a
nonlocal Higgs-type model. Due to the effect of stretching the potential
\cite{Lidsey,BarCline} nonlocality destroys the relation between the coupling
constant in the potential, the mass term and the value of the vacuum energy
(cosmological constant) and produces the effective double-well potential with
the negative extra cosmological term~\cite{ABG} (see details in Appendix).
Models with a non-minimal coupling
dilaton are being ~\cite{McGreevy:2005ci,Swanson:2008dt,Aref'eva:2008gj,EscamillaRivera:2011di} actively studied and are, in particular, motivated by the SFT.

In many scalar-driven inflationary scenarios the inflationary potential
 has a minimum and the scalar
field rolls to this minimum and starts oscillating around it.
Such behaviour of the scalar field allows us to avoid never-ending inflation.
In realistic models of inflation it is usually assumed that at the end of the
period of accelerated expansion (during the oscillatory regime) the scalar field
decays into ordinary matter, in other words, reheating takes place.
For a positive definite Higgs potential
we get convergent oscillations, whereas if we assume that the minimal
value of the potential is negative, then such process is not possible
because the value of the Hubble parameter should be real and there exists
a minimal non-zero velocity of the scalar field in the minimum.
The characteristic property of models with  non-positive definite potential
is the existence of the unreachable domain on the phase plane, which
corresponds to non-real Hubble parameter.
In the case of a minimal coupling, it turns out that  non-positivity of
a potential leads to an inevitable catastrophic change of the
Universe evolution: from expansion to contraction~\cite{ABG}. This is just an
opposite to what is going   in the cyclic model~\cite{Turok}, where the ekpyrotic
phase is generated by a scalar field
with a negative potential and  one has a bounce from a phase of
contraction to a phase of expansion, see~\cite{Steinhardt} for discussion of the
Planck2013 results support of this model.

In more details, the goal of the paper is to explore the behaviour
of the scalar field in the neighbourhood of the negative minimum of a potential
 in  modified gravity models.
We consider the modified gravity models
with action including the term proportional to the Ricci scalar multiplied
by a function of the scalar field and with a  negative minimal value of
double-well potential.
By numerical calculations we explore the induced gravity model and model, with
both the Hilbert--Einstein term and the squared scalar field multiplied by the
scalar curvature.
We also choose such values of the negative cosmological constant that the unreachable
domain is separated into the two parts. We show
that the phase trajectories are being attracted to the boundary of the unreachable domain,
touch it and go to infinity.
During this evolution the Hubble parameter oscillates from positive
to negative values.  This quasi-cyclic behaviour ends up on a point of the boundary
of the unreachable domain and after this the evolution changes and completes
by monotonically contracting stage.

The paper is organized as the following. We start, Section 2, from a short remind of the action and equations for the
non-minimally coupled cosmological models. We consider dynamics in the models of induced gravity and of the Higgs-like inflation with an non-positive definite double-well potential. We show that there are solutions that wind up  on the unreachable domain on the phase plane.   In Section 3, we discuss the model in the Jordan and Einstein frames. In Section 4, we study the special case
$\xi=-1/6$, in which some equations, considered in Section~2, have to be modified.
Finally, Section~5 is devoted to the conclusion.

\section{Classical dynamics in cosmological models with non-minimal coupling and non-positive definite potentials}
\subsection{Cosmological models with non-minimally coupled scalar fields}
Cosmological models in framework of a non-minimally coupled theory are being actively studied~\cite{nonmin-inf,Kaiser:1994,Kaiser:1994vs,HI,Bezrukov:2008ut,HI1,HI3,HI4,induced,Cooper:1982du,KKhT,Kamenshchik:2012rs,CervantesCota:2010cb,KTV,
Sami:2012uh,KTVV} (see also~\cite{Book-Capozziello-Faraoni,CL} and references therein).

Models with non-minimally coupled scalar fields are described by the following action:
\begin{equation}
\label{action}
S=\int d^4 x \sqrt{-g}\left[ U(\phi)R-\frac12g^{\mu\nu}\phi_{,\mu}\phi_{,\nu}-V(\phi)\right],
\end{equation}
where $U(\phi)$ and $V(\phi)$ are differentiable functions of the scalar field $\phi$.
We assume that $U(\phi)\geqslant0$.
We use the
signature $(-,+,+,+)$, and $g$ is the determinant of the metric tensor~$g_{\mu\nu}$.

In the spatially flat Friedmann--Lema\^{i}tre--Robertson--Walker (FLRW) metric with the interval:
\begin{equation}
\label{friedmangeneral}
ds^2={}-dt^2+a^2(t)\left(dx_1^2+dx_2^2+dx_3^2\,\right),
\end{equation}
we get the following equations
\begin{equation}
\label{Fr1}
6UH^2+6\dot UH = \frac{1}{2}{\dot\phi}^2
+ V,
\end{equation}
\begin{equation}\label{Fr2}
   2U\left(2\dot H+3H^2\right)={}-\frac{{\dot\phi}^2}{2}-2\ddot U-4H\dot U+ V.
\end{equation}
The equation for the field got by the variation of the action over the field is
\begin{equation}\label{Fieldequ}
    \ddot\phi+3H\dot\phi-6U'\left(\dot H+2H^2\right)+V'=0.
\end{equation}
here and hereafter a dot indicates the derivation over $t$, a prime indicates
the derivation over~$\phi$. The Hubble parameter is $H\equiv\dot a/a$.

Subtracting equation~(\ref{Fr1}) from equation~(\ref{Fr2}), we get
\begin{equation}\label{Fr21}
   4U\dot H={}-{\dot\phi}^2-2\ddot U+2H\dot U.
\end{equation}

From equations~(\ref{Fr1})--(\ref{Fr21}), we get the following system of the first order equations:
\begin{equation}
\begin{split}
\dot\phi&=\psi,\\
\dot\psi&={}-3H\psi-\frac{\left[(6 U''+1)\psi^2-4V\right]U'+2UV'}{2\left(3 {U'}^2+ U\right)},\\
\dot H&={}-\frac{2U''+1}{4(3{U'}^2+U)}\psi^2+\frac{2U'}{{3{U'}^2+U}}H\psi
-\frac{6{U'}^2}{3{U'}^2+U}H^2+\frac{U'V'}{2(3{U'}^2+U)}\,.
\end{split}
\label{FOSEQU}
\end{equation}

Note that equation~(\ref{Fr1}) is not a consequence of system  (\ref{FOSEQU}). On the other hand,
if equation~(\ref{Fr1}) is satisfied in the initial moment of time, then from system (\ref{FOSEQU}) it follows
that equation~(\ref{Fr1}) is satisfied at any moment of time. In other words,  system (\ref{FOSEQU}) is equivalent to the initial system of equations (\ref{Fr1})--(\ref{Fieldequ}) if and only if we choose such initial data that equation~(\ref{Fr1}) is satisfied.

Let us introduce a new variable
\begin{equation}
Q\equiv H+\frac{\dot U}{2U}.
\end{equation}
In terms of $Q$ equations~(\ref{Fr1}) and (\ref{Fr21}) have the following form
\begin{equation}
3Q^2=\frac{{\dot\phi}^2}{4U}+\frac{3{\dot U}^2}{4U^2}+\frac{V}{2U}.
\end{equation}
\begin{equation}
\label{Fr21Q}
\dot Q-\frac{\dot U}{2U}Q={}-\frac{U+3{U'}^2}{4U^2}\,{\dot\phi}^2.
\end{equation}
Therefore,
\begin{equation}
\label{Fr21Qm}
\frac{d}{dt}\left[\frac{Q}{\sqrt{U}}\right]={}-\frac{U+3{U'}^2}{4U^2\sqrt{U}}\,{\dot\phi}^2 \leqslant 0.
\end{equation}
For an arbitrary positive function $U(\phi)$ we get that $Q/\sqrt{U}$ is monotonically decreasing function. If for some moments of time $t_1$ and $t_2>t_1$ we have $\phi(t_2)=\phi(t_1)$ and $\phi(t)$ is not a constant at $t_1\leqslant t\leqslant t_2$, then $Q(t_2)<Q(t_1)$.
The physical reasons of inequality (\ref{Fr21Qm}) will be clear in the next section, where we consider this model in the Einstein frame.

All above-mentioned formulae are valid for an arbitrary potential $V(\phi)$. Let us consider a non-positive definite potential. In this case,
on the $(\phi,\dot{\phi})$ plane there is the boundary, at any point of which $Q=0$, that is equivalent to
\begin{equation}
\label{unr-boundary}
\frac{{\dot\phi}^2}{2}+\frac{3 (U'\,\dot\phi )^2}{2U}+V=0\,.
\end{equation}
Note that this boundary divides the phase plane into two domains: one corresponds to real values of the Hubble parameter, defined by (\ref{Fr1}), the other one corresponds to non-real values of this function.
So, if a trajectory starts from the real value of $H$, then it never crosses the line $Q=0$, but can touch this line.

We will call the domain on the $(\phi,\dot{\phi})$ plane, which corresponds
to non-real values of the Hubble parameter, defined by (\ref{Fr1}), as "unreachable domain", because
this domain is unreachable for solutions with the initial conditions, which satisfy (\ref{Fr1}) and correspond to real values of the Hubble parameter~$H$.
The boundary of this domain is defined by (\ref{unr-boundary}).

\subsection{Induced gravity models}
Let us consider the induced gravity models~\cite{Kaiser:1994vs,KTV} with
\be
U(\phi)= \frac12 \xi \phi^2\,,
\ee
where $\xi$ is the non-minimal coupling constant.

Equations (\ref{Fr1})--(\ref{Fieldequ}) for such a choice of the function $U(\phi)$ look as follows:
\be
\label{e2}
H^2=\frac{V}{3\xi\phi^2}+\frac{1}{6\xi}\left(\frac{\dot{\phi}}{\phi}\right)^2-2H\frac{\dot{\phi}}{\phi},
\ee
\be
\label{Equ11}
3H^2+2\dot{H}={}-2 \frac{\ddot{\phi}}{\phi}-4H\frac{\dot{\phi}}{\phi}
-\frac{4\xi+1}{2\xi}\left(\frac{\dot{\phi}}{\phi}\right)^2+ \frac{V}{\xi\phi^2},
\ee
\be
\label{Equ_phi}
\ddot{\phi}+3H\dot{\phi}+V'-6\xi\phi\left(2H^2+\dot{H}\right)=0.
\ee

We assume that $\xi\neq -1/6$. The case $\xi= -1/6$ will be considered separately in Section~\ref{xi_16}.
All calculations have been made for $\xi>0$ that corresponds to $U(\phi)\geqslant 0$.

Equation (\ref{e2}) is the quadratic equation for $H$:
\be\label{FV}
H^2+2H\frac{\psi}{\phi}-\frac{V}{3\xi\phi^2}-\frac{1}{6\xi}\left(\frac{\psi}{\phi}\right)^2=0,
\ee
and has the following solutions
\be\label{HV}
H_\pm={}-\frac{\psi}{\phi}\pm\sqrt{\left(1+\frac{1}{6\xi}\right)\left(\frac{\psi}{\phi}\right)^2
+\frac{V}{3\xi\phi^2}}.
\ee

The function $H$ is a continuous function, so, if $V(\phi)>0$ for all $\phi$, then evolution of the Universe in such a model is described either only $H_-$ or only $H_+$. It depends on initial conditions. If $V(\phi)$ is not a positive definite function, then it is possible that a part of evolution is described by $H_-$, whereas the other part --- by $H_+$. The simplest way to get a non-positive definite potential from the known positive definite one is to subtract a positive constant.

For the induced gravity model
\begin{equation*}
Q\equiv H+\frac{\psi}{\phi}
\end{equation*}
and equation~(\ref{Fr21Q}) has the following form
\begin{equation}
\label{equq_Q}
    \dot Q=\frac{\psi}{\phi}Q-\frac{6 \xi+1}{2\xi} \left(\frac{\psi}{\phi}\right)^2.
\end{equation}
It is easy to see that for positive $\xi$ and $\phi$
\begin{equation}\label{dQ_phi}
    \frac{d}{dt}\left[\frac{Q}{\phi}\right]={}-\frac{6 \xi+1}{2\xi\phi} \left(\frac{\psi}{\phi}\right)^2\leqslant 0.
\end{equation}

Combining equations (\ref{e2})--(\ref{Equ_phi}),  we  obtain the following system of the first order differential equations
\be
\label{2}
\dot{\phi}=\psi\,,
\ee
\be
\dot{\psi}={}-3H\psi-\frac{\psi^2}{\phi}+\frac{1}{(1+6\xi)\phi}[4V(\phi)-\phi V'(\phi)],
\label{e3}
\ee
\be
\label{3}
\dot{H}=\frac{4H\psi}{(1+6\xi)\phi} +\frac{V'(\phi)}{(1+6\xi)\phi}-\frac{12\xi}{1+6\xi}H^2-\frac{1+2\xi}{2\xi(1+6\xi)}\left(\frac{\psi}{\phi}\right)^2. \ee
Equation~(\ref{FV}) is a consequence of system (\ref{2})--(\ref{3}), under condition that the initial values of the functions $\phi$, $\psi$ and $H$ satisfy~(\ref{FV}).

From equations (\ref{e2}) and (\ref{Equ11}) we get the following equation, whose form does not depend on the potential:
\begin{equation}
\label{Equ12}
\dot{H}={}- \frac{\dot{\psi}}{\phi}+H\frac{\psi}{\phi}-\frac{2\xi+1}{2\xi}\left(\frac{\psi}{\phi}\right)^2.
\end{equation}
Substituting the value of $H$ on the boundary of the unreachable domain, we get the following equation
\begin{equation}
\frac{6\xi+1}{\xi}\left(\frac{\psi}{\phi}\right)^2=0.
\end{equation}
It means that the boundary of the unreachable domain is not a solution for system (\ref{2})--(\ref{3}) at any values of parameters.
Only trivial solutions with $\psi=0$ can belong to this boundary. The above-mentioned results are valid for any non-positive definite potential.

Models of this type in the case of the  Higgs potential have been considered in a lot of works (see, for example, \cite{HI,Bezrukov:2008ut,induced,HI1,HI3}).
In this paper, dynamics in the model with the following non-positive definite Higgs-like potential
\be
\label{V_H}
 V_H(\phi)=\frac{\varepsilon}{4}\left(\phi^2-b^2\right)^2-\Lambda,\qquad \Lambda>0,
\ee
where $\varepsilon$, $b$ and $\Lambda$ are  constants, is studied.

In the case of the $V_H(\phi)$ potential, the boundary of the unreachable domain ($Q=0$)  has the following form
\be
(1+6\xi)\dot{\phi}^2=2\Lambda-\frac{\varepsilon}{2}\left(\phi^2-b^2\right)^2.
\ee
Most attention will be paid to the case when $\Lambda<\varepsilon b^4/4$. In this case, the unreachable domain on the $(\phi,\dot{\phi})$ plane consists of two separate parts.

\begin{figure}[!h]
\centering
\includegraphics[width=67mm]{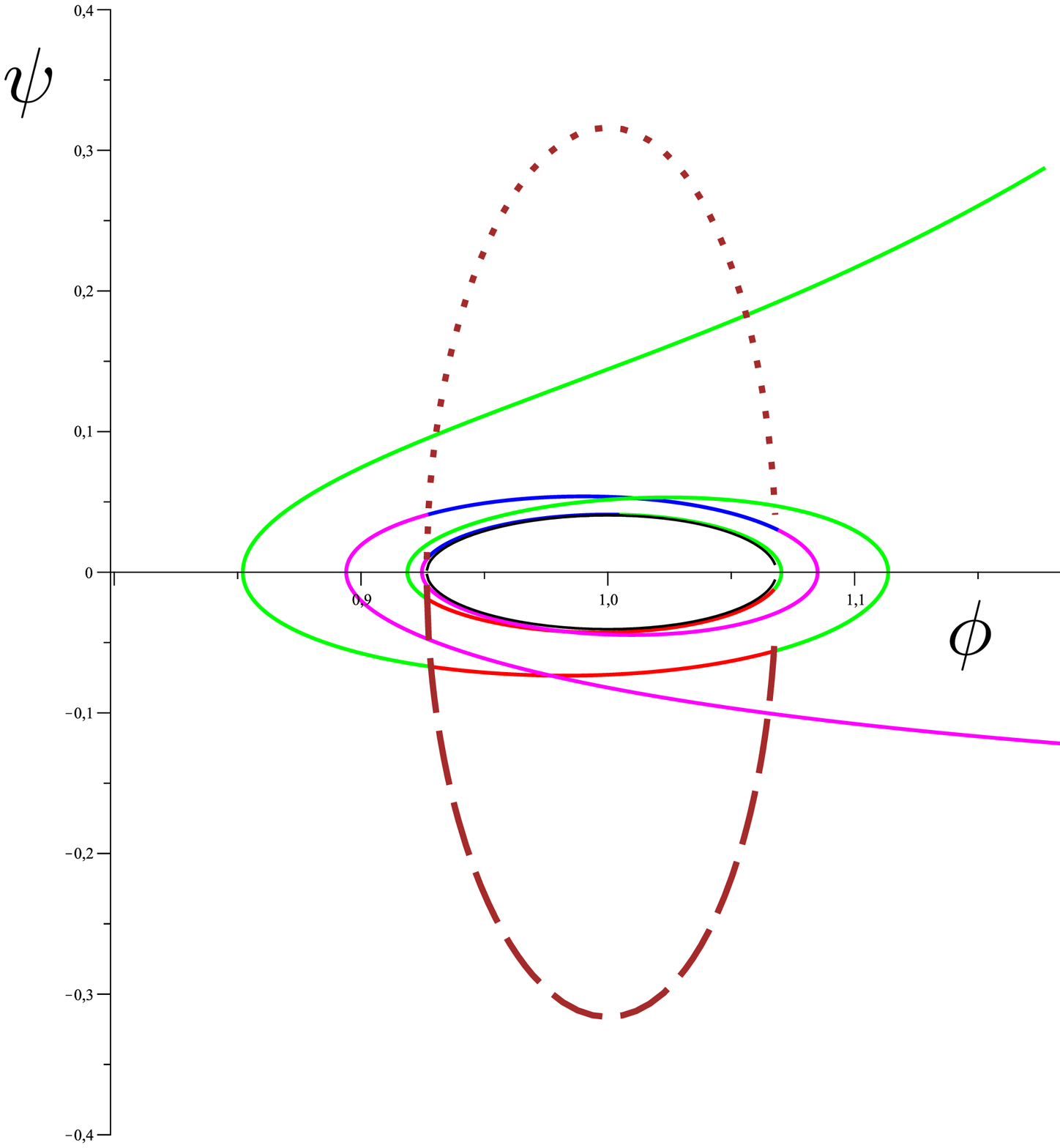} \  \  \  \  \  \  \
\includegraphics[width=67mm]{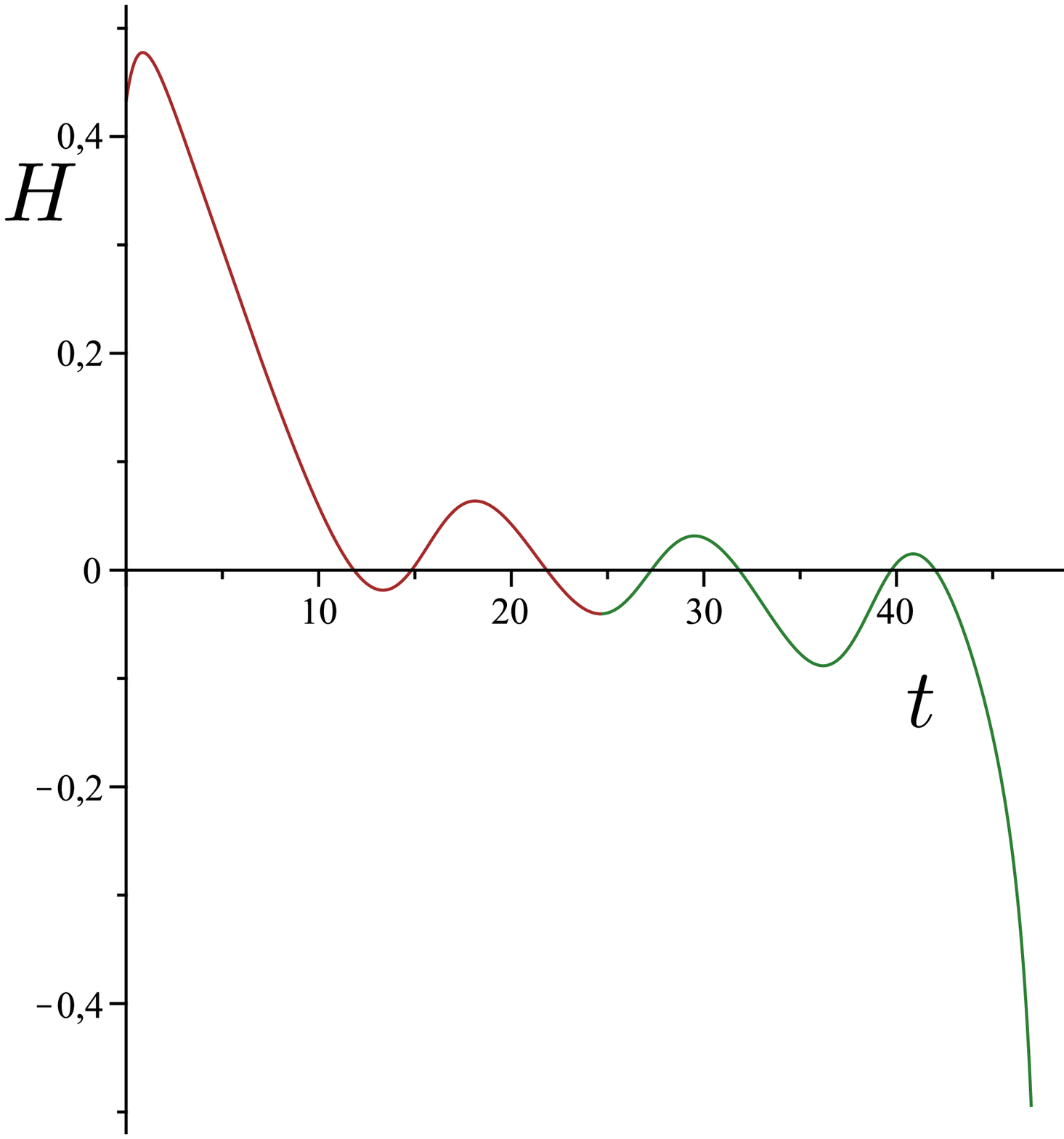}
\caption{The solution of system (\ref{2})--(\ref{3}) at $\Lambda=0.05$. We choose $b=1$, $\epsilon=10$, $\xi=10$. The initial conditions are $\phi_0=2$, $\psi_0=0$, $H_0$ is calculated by (\ref{HV}) with sing "+". On the left picture, the phase diagram is present. It is magenta at the corresponding value of $H=H_+>0$, blue at $H=H_+<0$, red at $H=H_->0$, and green at $H=H_-<0$ brown. On this picture, brown dashed line corresponds to $H_+=0$, brown dashed line with long dashes corresponds to $H_-=0$, black line is the boundary of the unreachable domain.  On the right picture, the Hubble parameter as function of the cosmic time is present. Brown color means that $H=H_+$, whereas $H=H_-$ is drown in dark green color.}
\label{Psi_Phi_H}
\end{figure}

Numerical calculations give the following phase trajectory for system (\ref{2})--(\ref{3}).
As we see there are two stages of evolution on the phase plane. The first stage corresponds to some kind of the cyclic Universe, then the phase trajectory reaches the boundary of the unreachable domain and the second stage starts, during this stage the Hubble parameter rapidly decreases and the Universe contracts (See~Fig.\ref{Psi_Phi_H}).

The fact which gives the change of the stages is that when the phase trajectory reaches the boundary of the unreachable domain the square root in the equation for the Hubble parameter gets the "minus" sign.

\begin{figure}[!h]
\centering
\includegraphics[width=70mm]{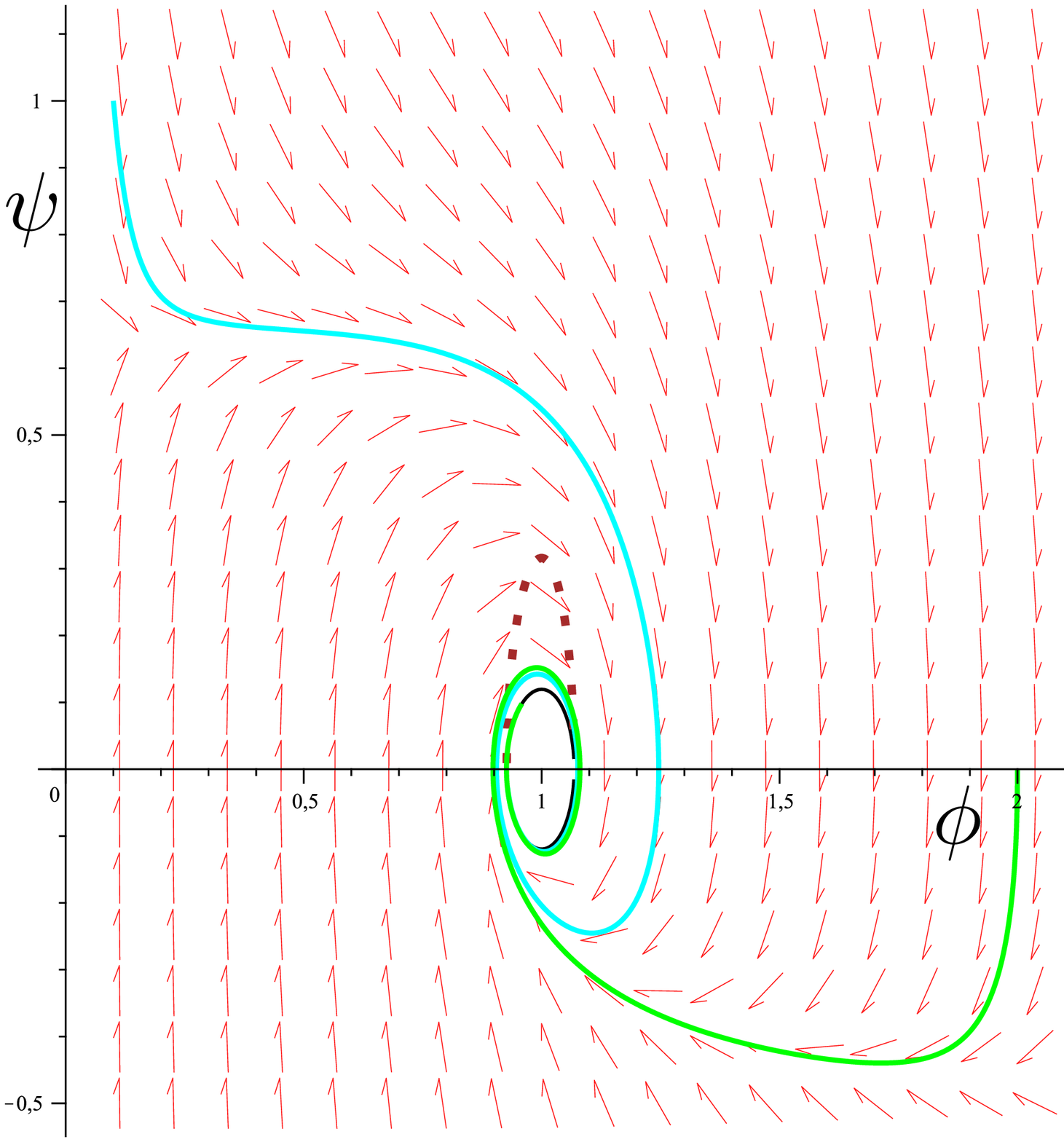} \  \  \  \  \ \  \
\includegraphics[width=57mm]{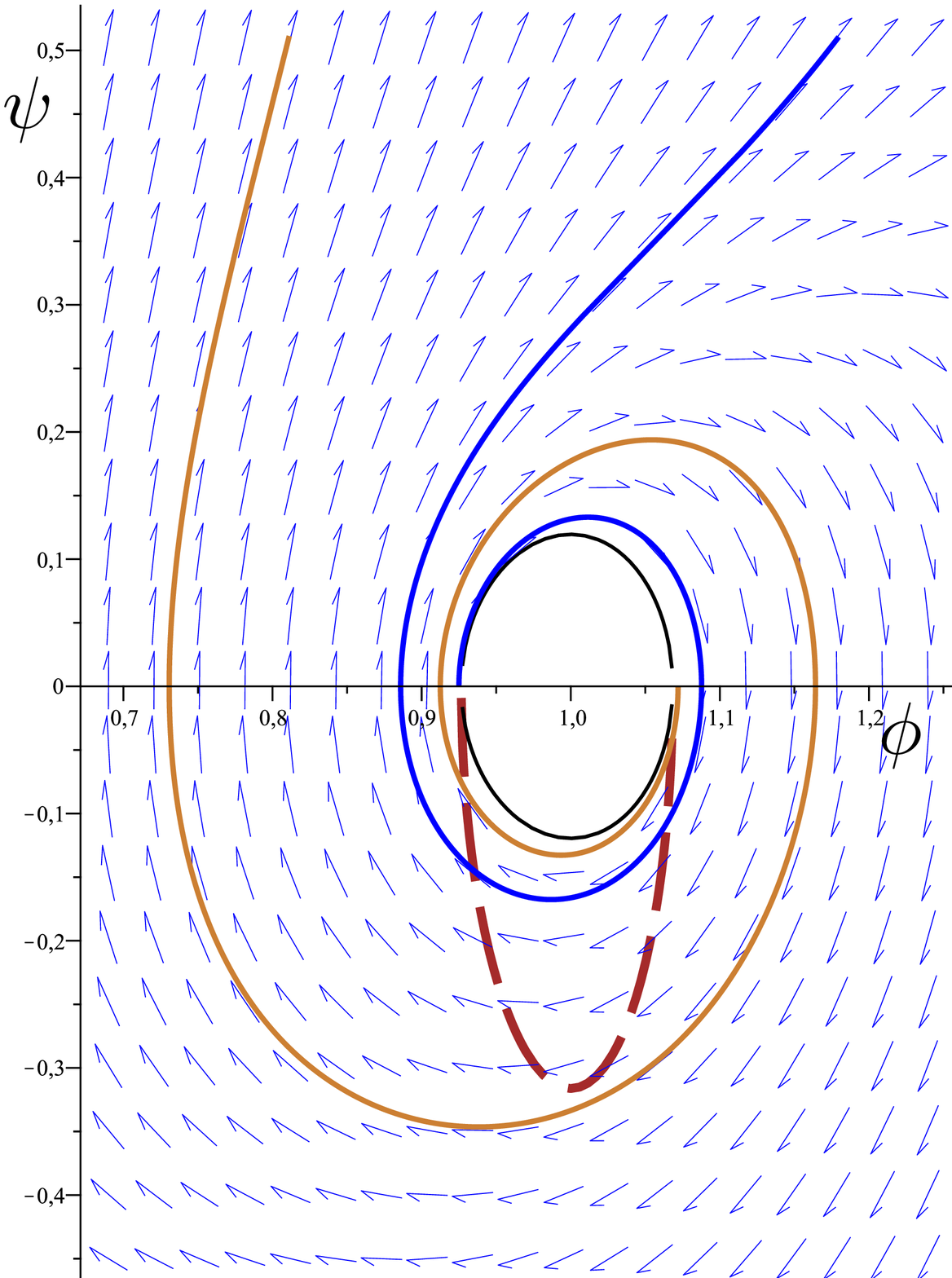}
\caption{Solutions at phase portraits of system (\ref{2})--(\ref{e3}) at $\Lambda=0.05$. We choose $b=1$, $\epsilon=10$, $\xi=1$. The function $H(t)$ is given by (\ref{FV}) as $H_+$ on the left picture and $H_-$ on the right picture. The initial conditions for the trajectories are $\phi_0=0.1$, $\psi_0=1$, $H=H_+$ (left) and $\phi_0=0.92$, $\psi_0=0$, $H=H_-$ (right).
On these pictures, brown dashed line corresponds to $H=0$, black line is the boundary, which corresponds to $Q=0$.}
\label{Phase_portraits}
\end{figure}

We can describe this stage of the evolution in more details. The evolution of the system (\ref{2})--(\ref{3}) is the following. Let us start with $\phi>0$, some $\psi$ and the corresponding $H=H_+$. The phase portrait on Fig.~\ref{Phase_portraits} (left picture) shows that the boundary $Q=0$ is attractive. So, the trajectory are winding up this boundary and then, at some finite moment of time,
touches it. From (\ref{equq_Q}), we get that always $\dot Q\leqslant 0$ at $Q=0$ and $\dot Q < 0$ at $Q=0$ and $\psi\neq 0$. This means that if a solution for system (\ref{2})--(\ref{3})  with $H_+$ touches this boundary at $\psi\neq 0$, then it changes to $H_-$, solutions with $H_-$ can not touch this boundary at $\psi\neq 0$. After this the solution tends to infinity (see Fig.~\ref{Phase_portraits}, right picture) and $H$ tends to $-\infty$.

We can address the following question: is there a possibility that the process of reaching the boundary of the unreachable domain will take the infinite time and the evolution won't finish by the contraction? The answer is that the process of reaching the boundary of the unreachable domain in the case when the trajectory rotates around it will always take a finite time\footnote{This conclusion doesn't depend on a specific form of the potential $V(\phi)$. The important property of this potential is that its minimum can not be reached with infinitely small velocity.}.

Indeed, if for some moments of time $t_1$ and $t_2>t_1$ we have $\phi(t_2)=\phi(t_1)$, then we get from (\ref{dQ_phi}) we get
\begin{equation*}  \frac{Q(t_2)}{\phi(t_2)}-\frac{Q(t_1)}{\phi(t_1)}=\frac{1}{\phi(t_1)}(Q(t_2)-Q(t_1))
={}
-\frac{6 \xi+1}{2\xi}\!\int\limits_{t_1}^{t_2}\frac{\psi^2}{\phi^3}\,dt\leqslant C_0<0.
\end{equation*}
where $C_0$ is a negative number.
So, for any circle value of $Q$ decreases on some positive value, which doesn't tend to zero, when number of circles tends to infinity, hence, only a finite number of circles is necessary to get the value $Q=0$.

Note that, when $Q$ tends to zero,
\begin{equation*}
\dot Q \rightarrow -\frac{6 \xi+1}{2\xi} \left(\frac{\psi}{\phi}\right)^2.
\end{equation*}

From (\ref{HV}) we get that $H$ is a real number at $\phi=b$,
only if
\begin{equation}
\left(1+6\xi\right)\psi^2>2\Lambda.
\end{equation}

Therefore, at $\phi\approx b$ and $Q\rightarrow 0$ we get
\begin{equation}
\dot Q \approx -\frac{6 \xi+1}{2\xi} \left(\frac{\psi}{b}\right)^2 < -\frac{\Lambda}{\xi b^2}.
\end{equation}
We come to conclusion that $\dot Q$ does not tend to zero at $Q\rightarrow 0$ if $\Lambda>0$.

\begin{figure}[!h]
\centering
$\varepsilon=4$\\[2.7mm]
\includegraphics[width=37mm]{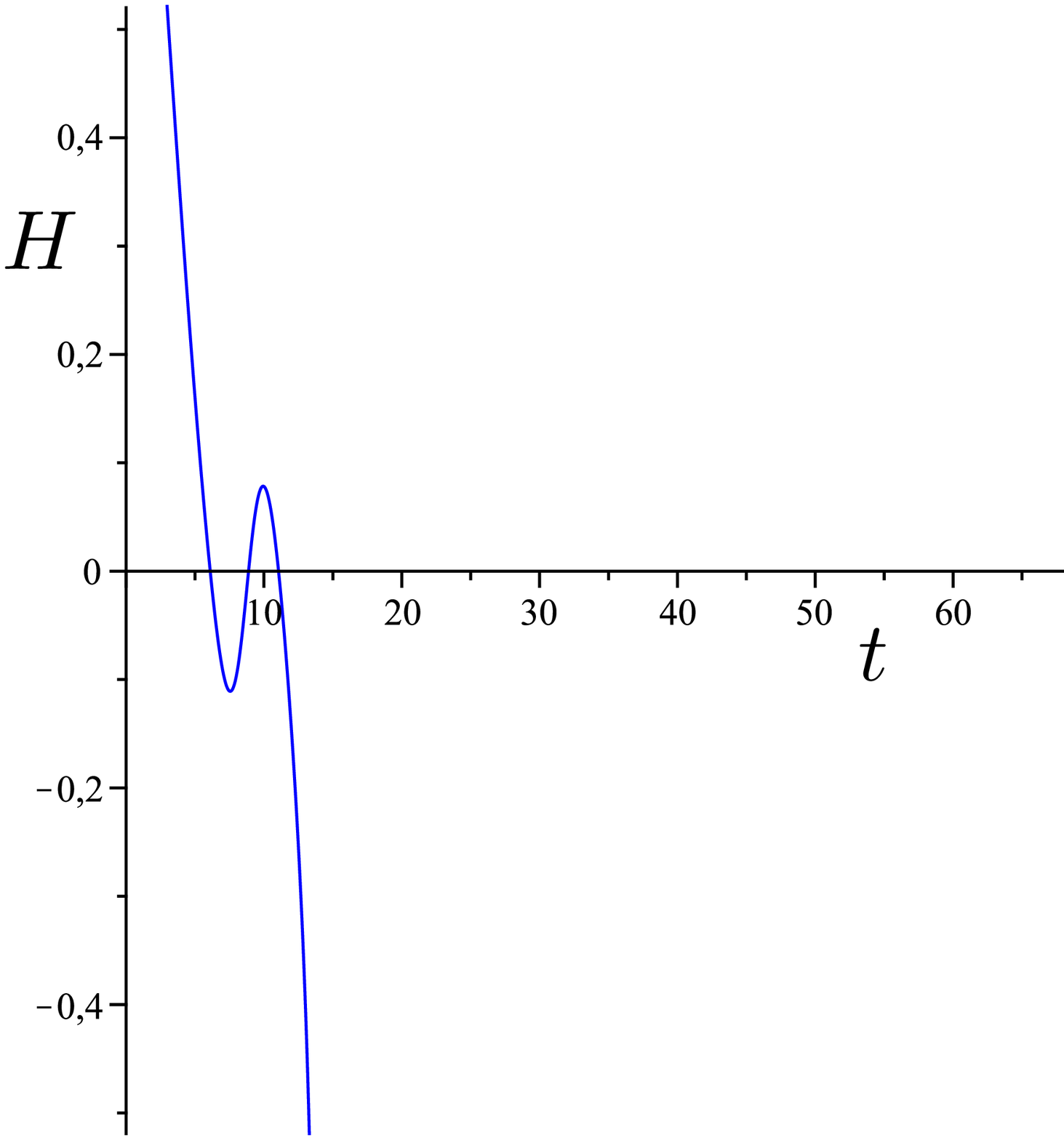}\  \  \  \  \
\includegraphics[width=37mm]{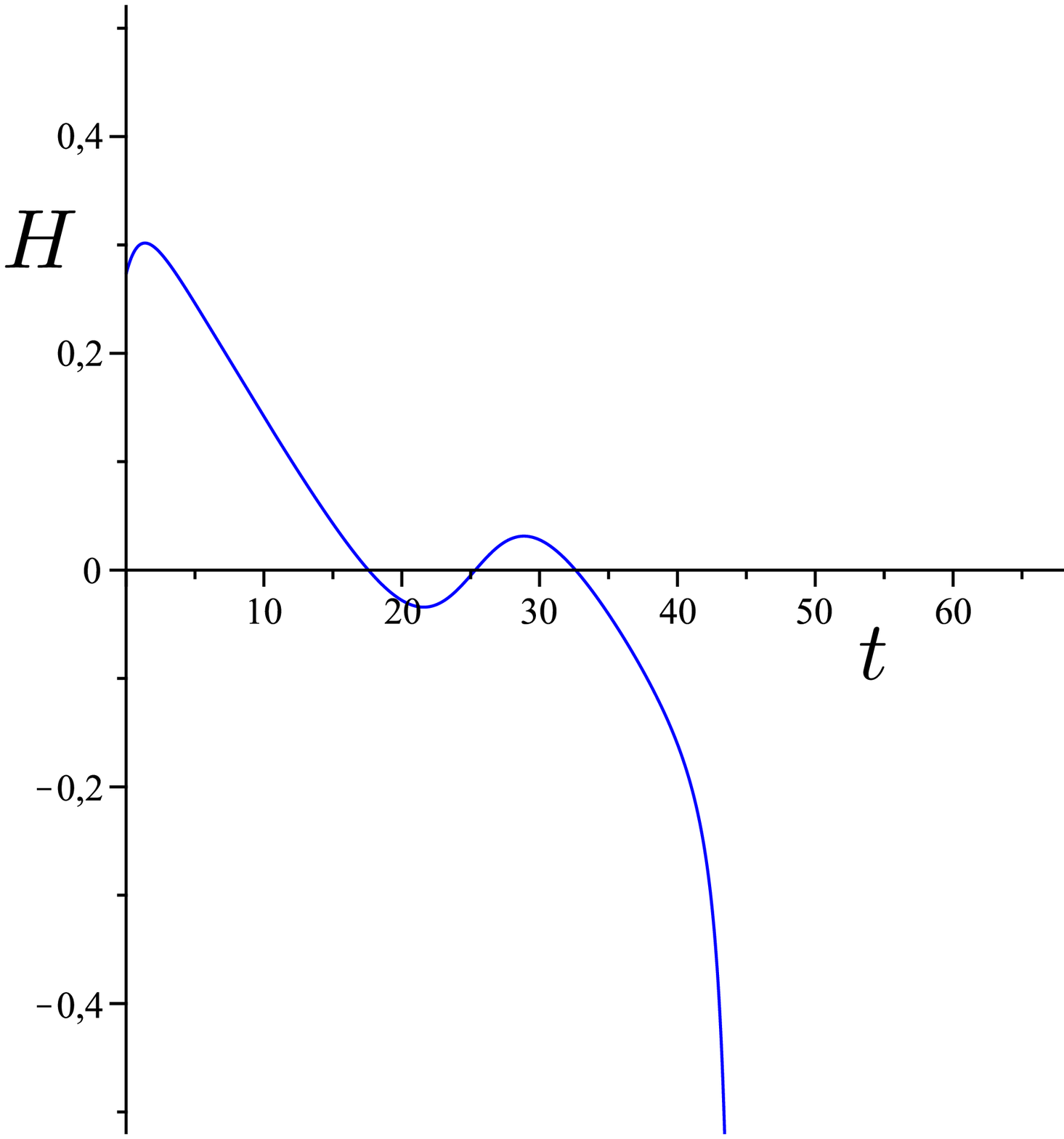}\  \  \  \  \
\includegraphics[width=37mm]{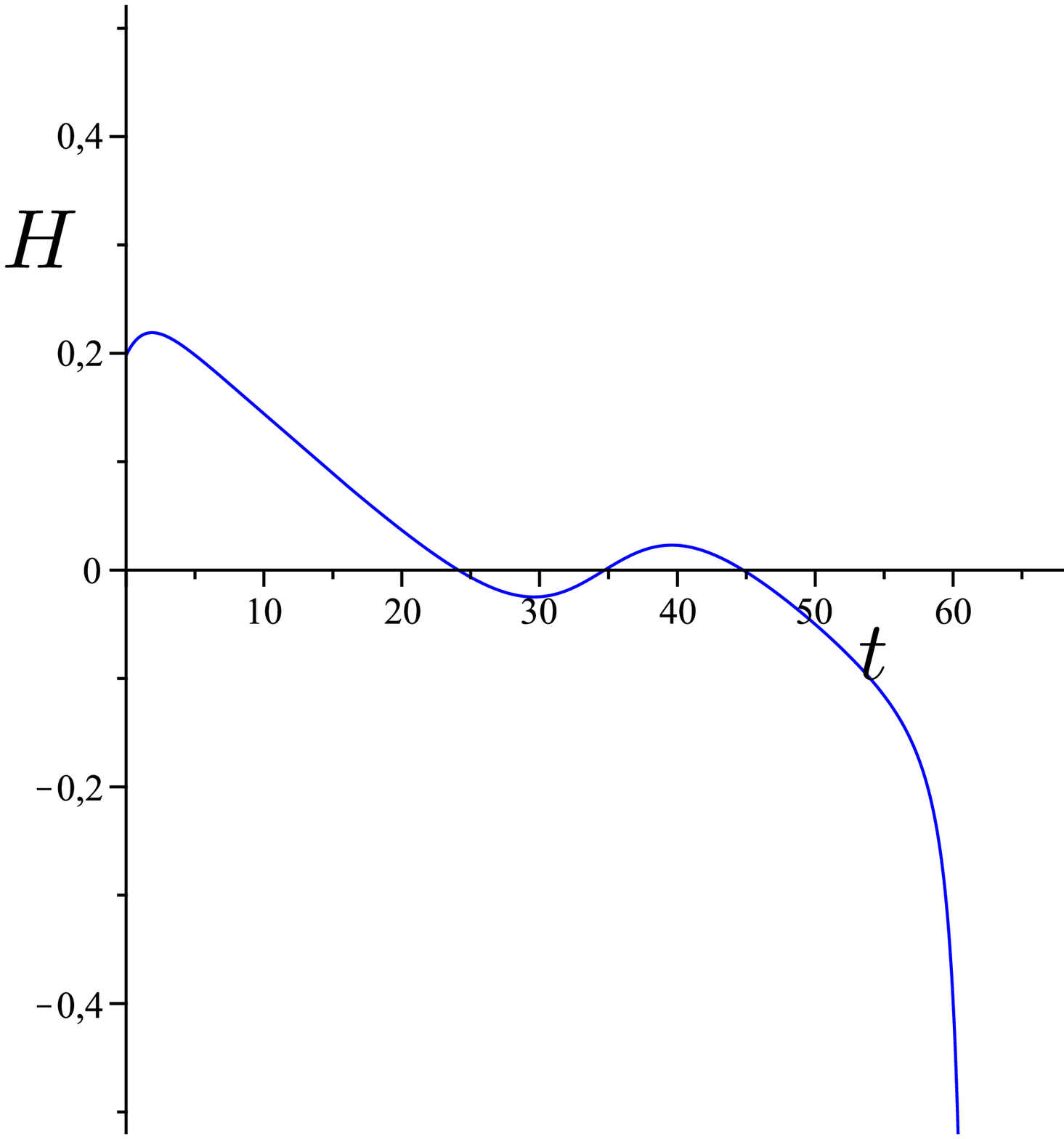}\\
$\varepsilon=10$\\[2.7mm]
\includegraphics[width=37mm]{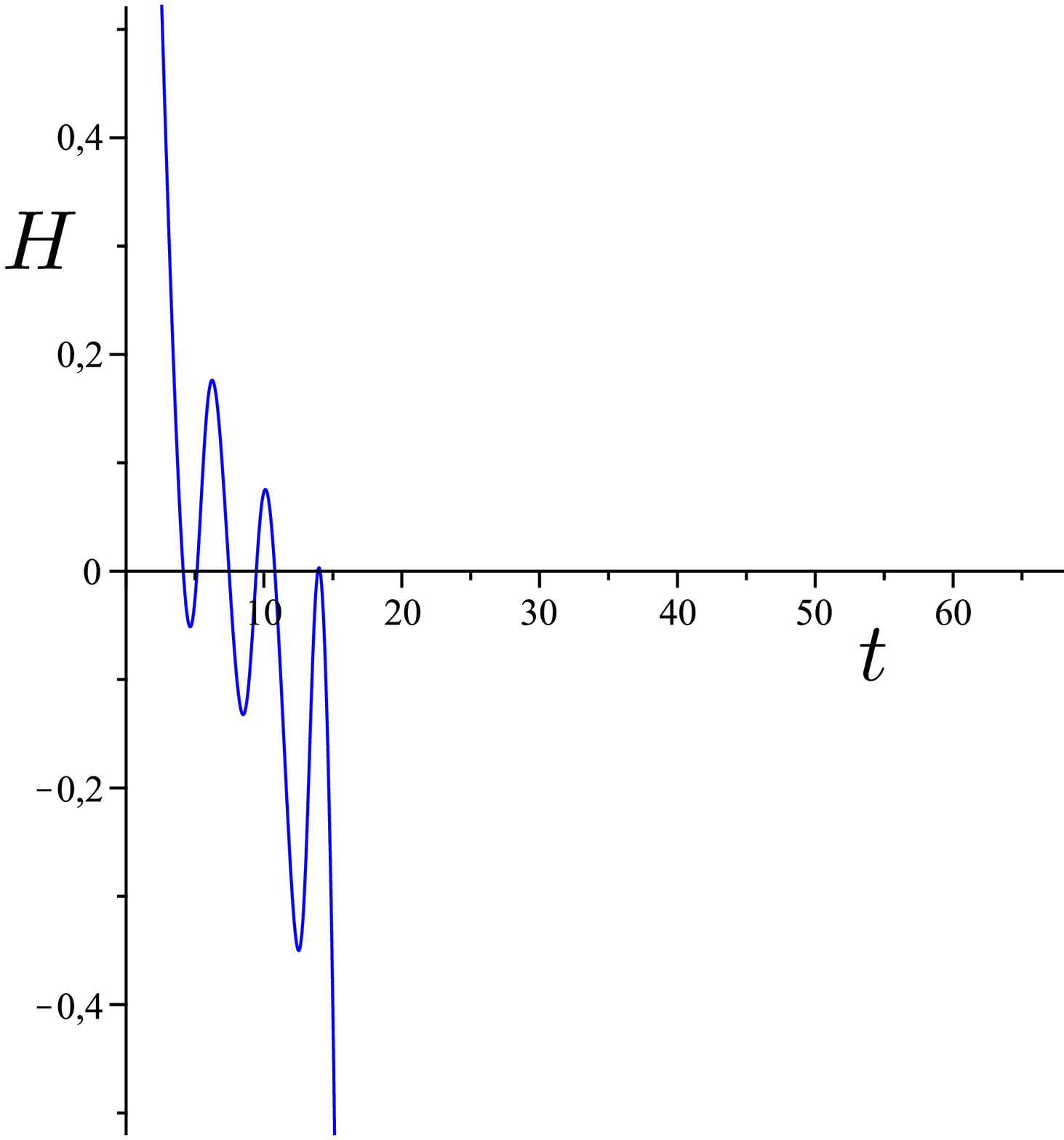}\  \  \  \  \
\includegraphics[width=37mm]{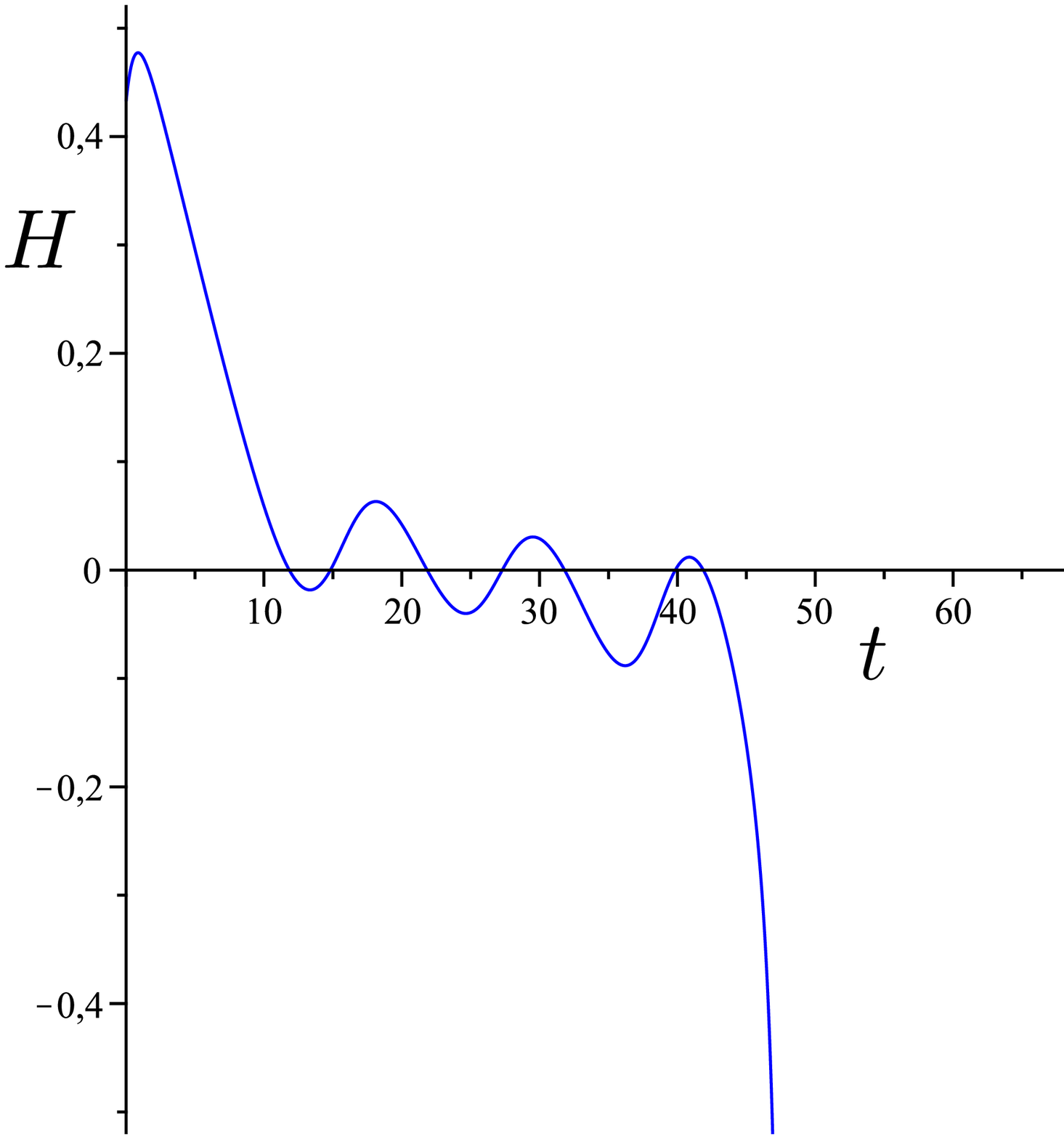}\  \  \  \  \
\includegraphics[width=37mm]{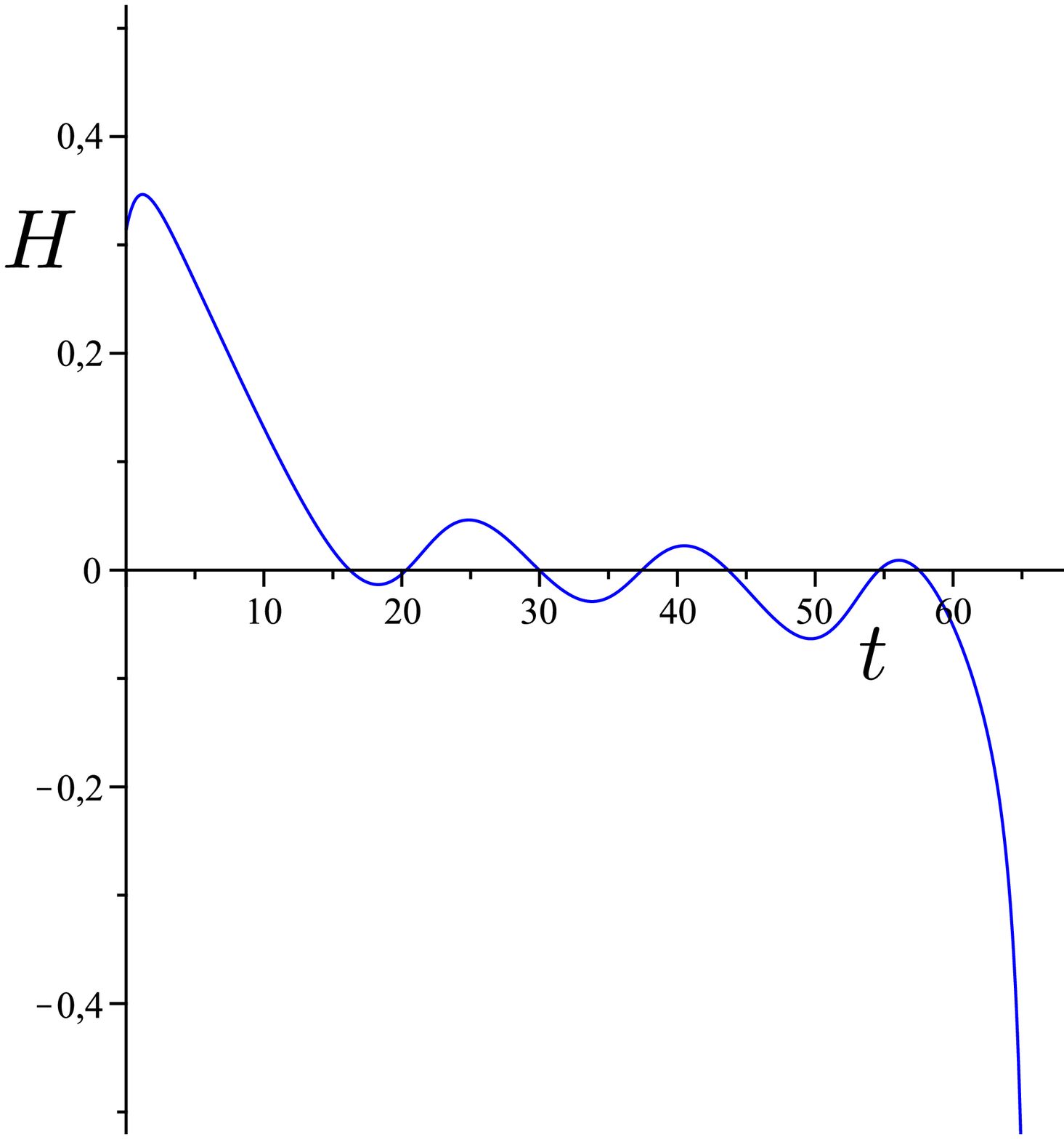}\\
$\varepsilon=100$\\[2.7mm]
\includegraphics[width=37mm]{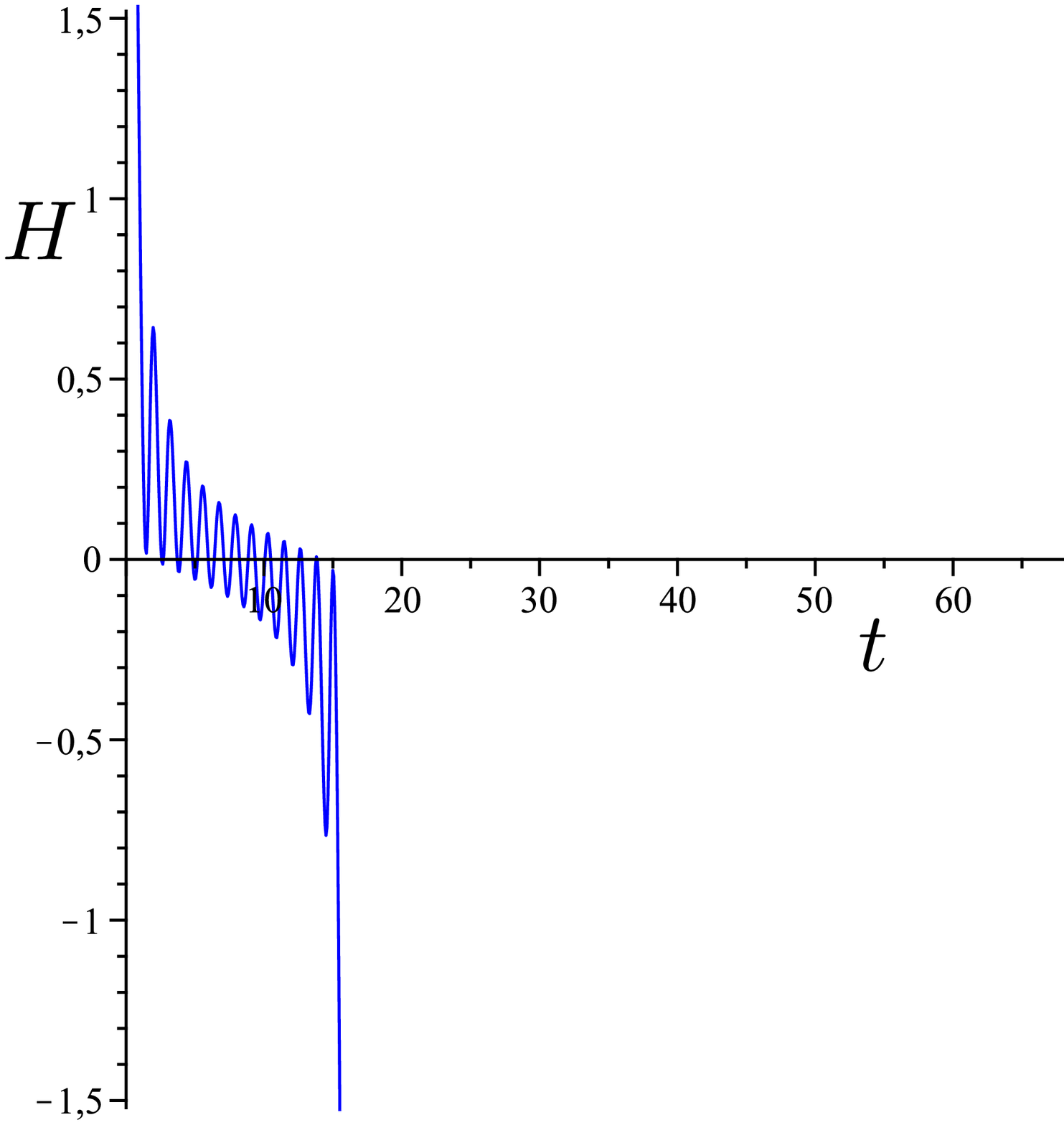}\  \  \  \  \
\includegraphics[width=37mm]{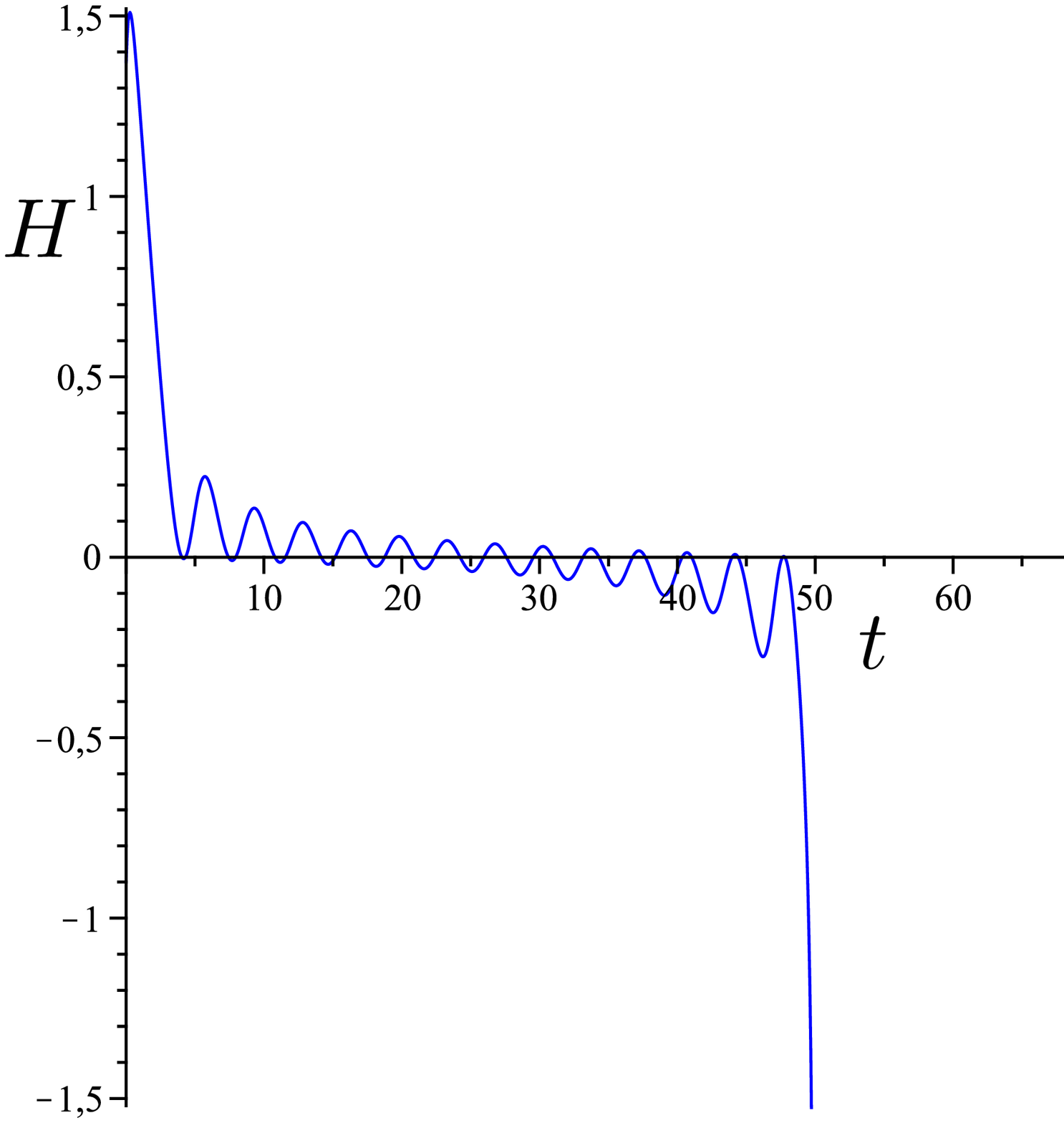}\  \  \  \  \
\includegraphics[width=37mm]{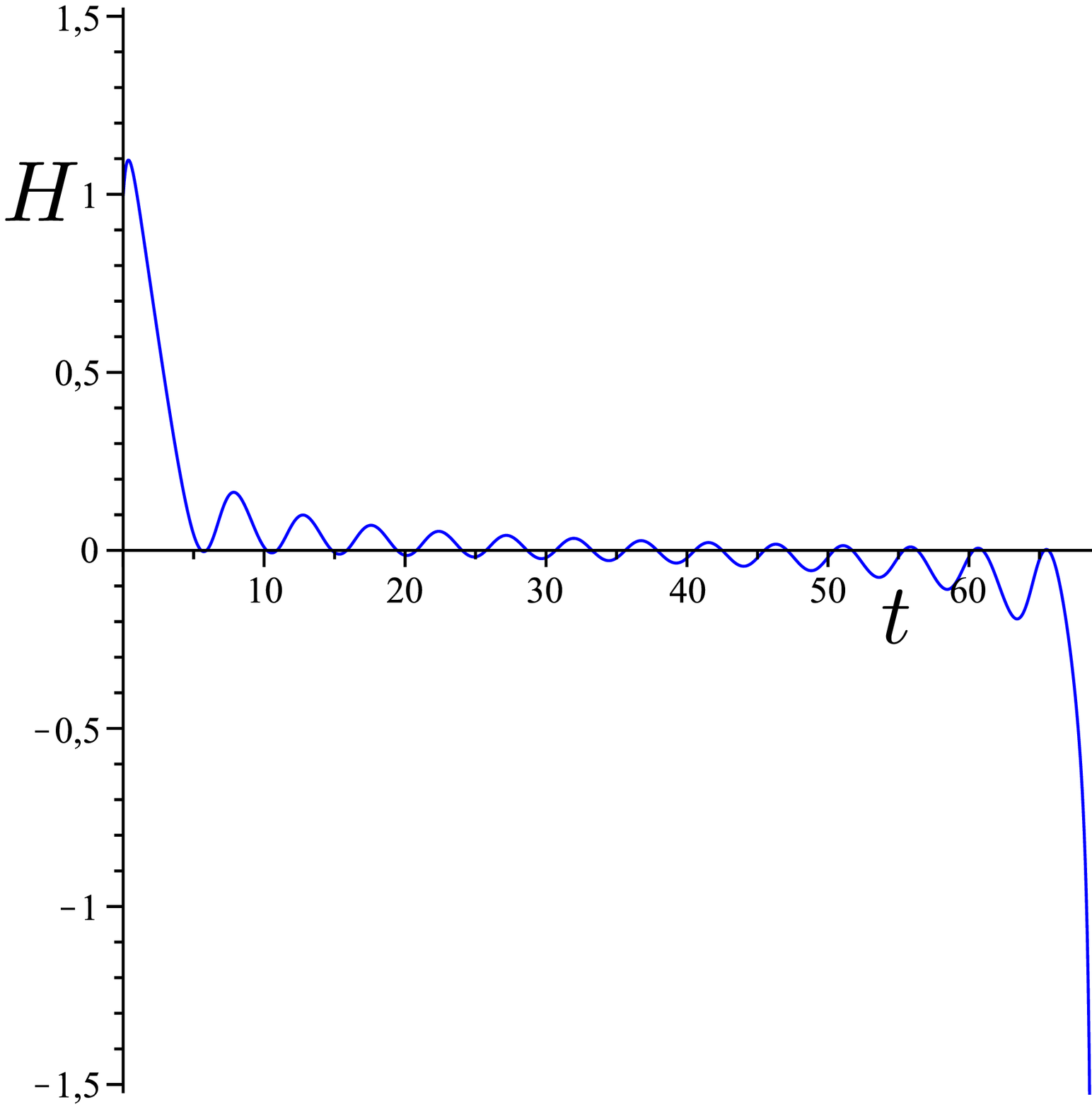}
\caption{The behaviour of the Hubble parameter at different values of $\xi$ and $\varepsilon$. We choose $b=1$ and $\Lambda=0.05$. The initial conditions are $\phi_0=2$, $\psi_0=0$, $H_0$ is calculated by (\ref{HV}) with sing "+". The parameter $\xi=1$ for the left column, $\xi=10$ for the left column, and $\xi=19$ for the right column. The parameter $\varepsilon=4$ for the first line, $\varepsilon=10$ for the second line, $\varepsilon=100$ for the third line.  }
\label{H_t_xi_eps}
\end{figure}

Let us analyse how the behaviour of the Hubble parameter depends on values of parameters $\xi$ and $\varepsilon$.
It is easy to see (Fig.~\ref{H_t_xi_eps}) that the number of the Hubble parameter oscillations increases with the parameter $\epsilon$, whereas a period of the oscillations increases with the parameter $\xi$.

\begin{figure}[!h]
\centering
\includegraphics[width=55mm]{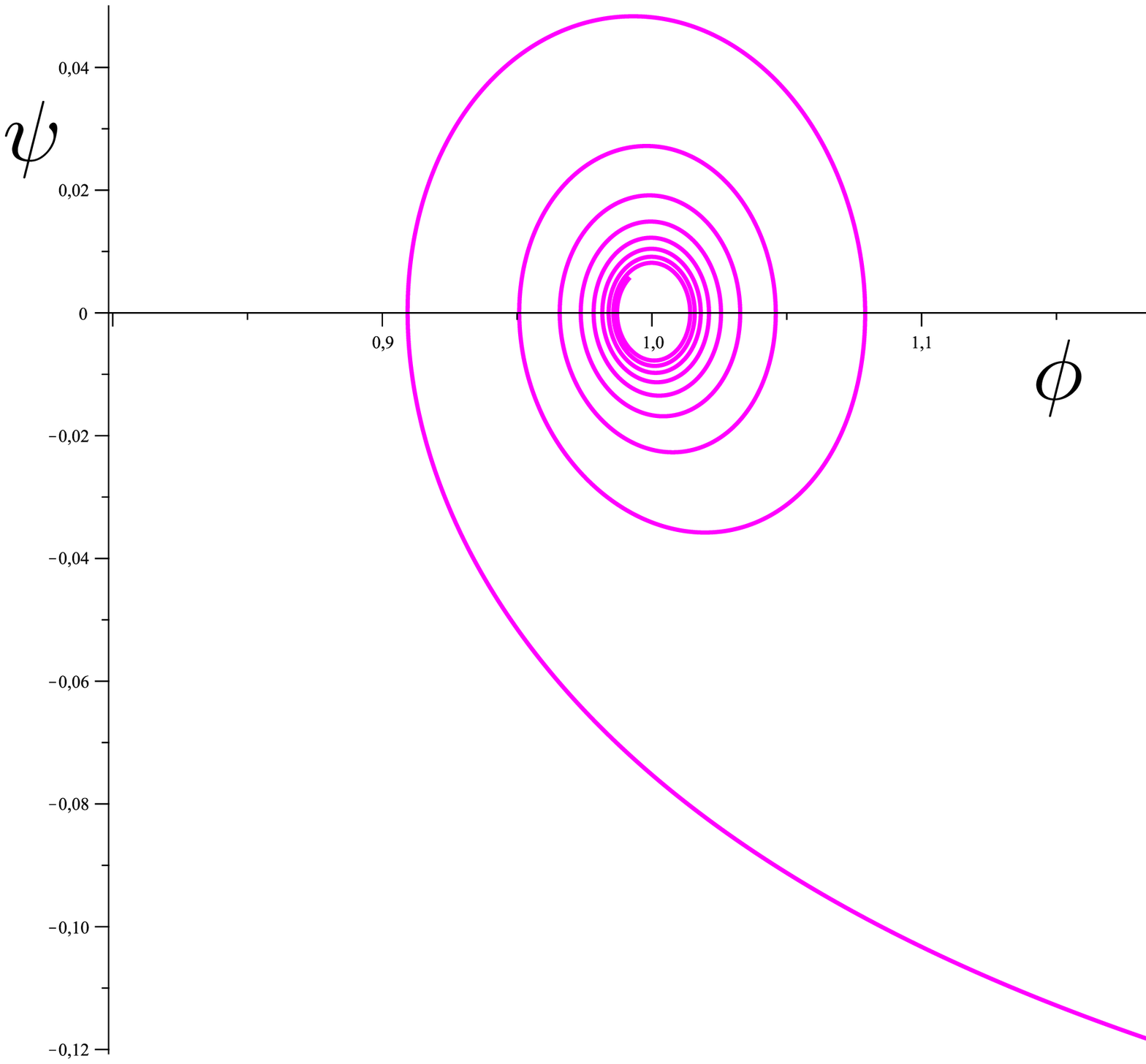}\  \  \  \  \  \  \  \  \
\includegraphics[width=55mm]{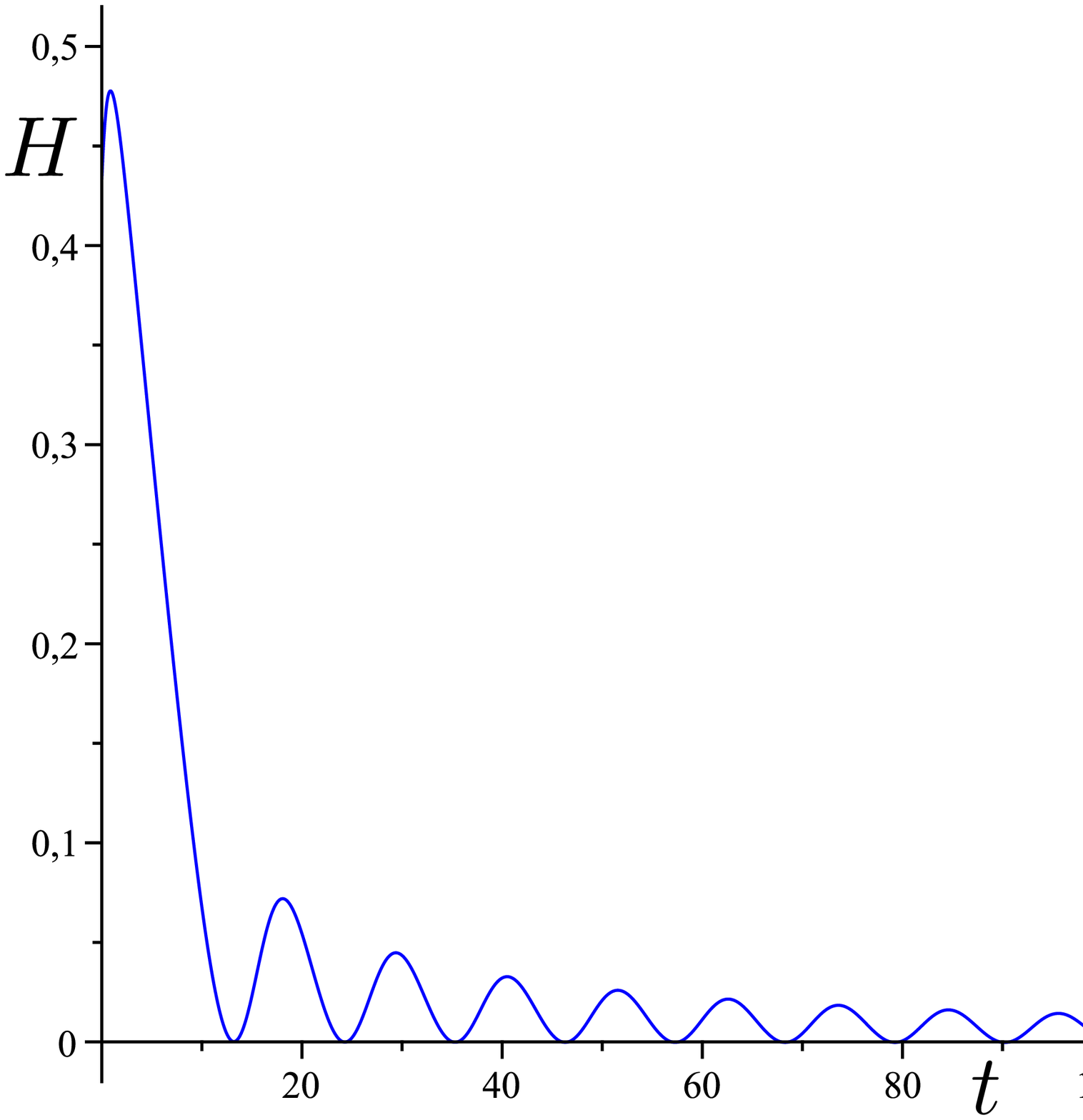}
\caption{The solution of system (\ref{2})--(\ref{3}) at $\Lambda=0$. We choose $b=1$, $\varepsilon=10$, $\xi=10$. The initial conditions are $\phi_0=2$, $\psi_0=0$, $H_0$ is calculated by (\ref{HV}) with sing "+" ($H_0=\sqrt{3}/4$). The Hubble parameter is always $H_+$.}
\label{PhPlLambda0}
\end{figure}

\begin{figure}[!h]
\centering
\includegraphics[width=55mm]{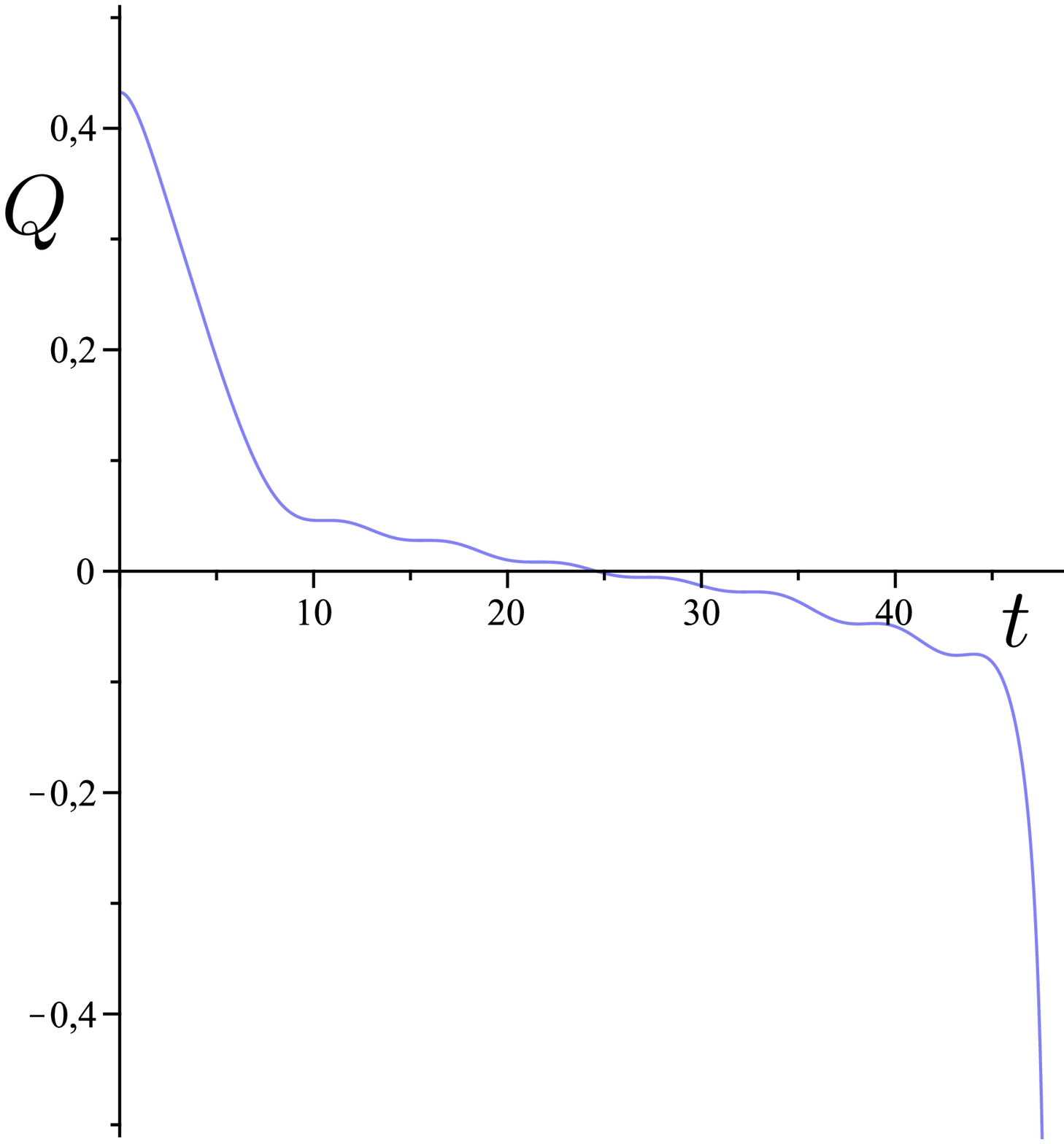}\  \  \  \  \  \  \  \  \
\includegraphics[width=55mm]{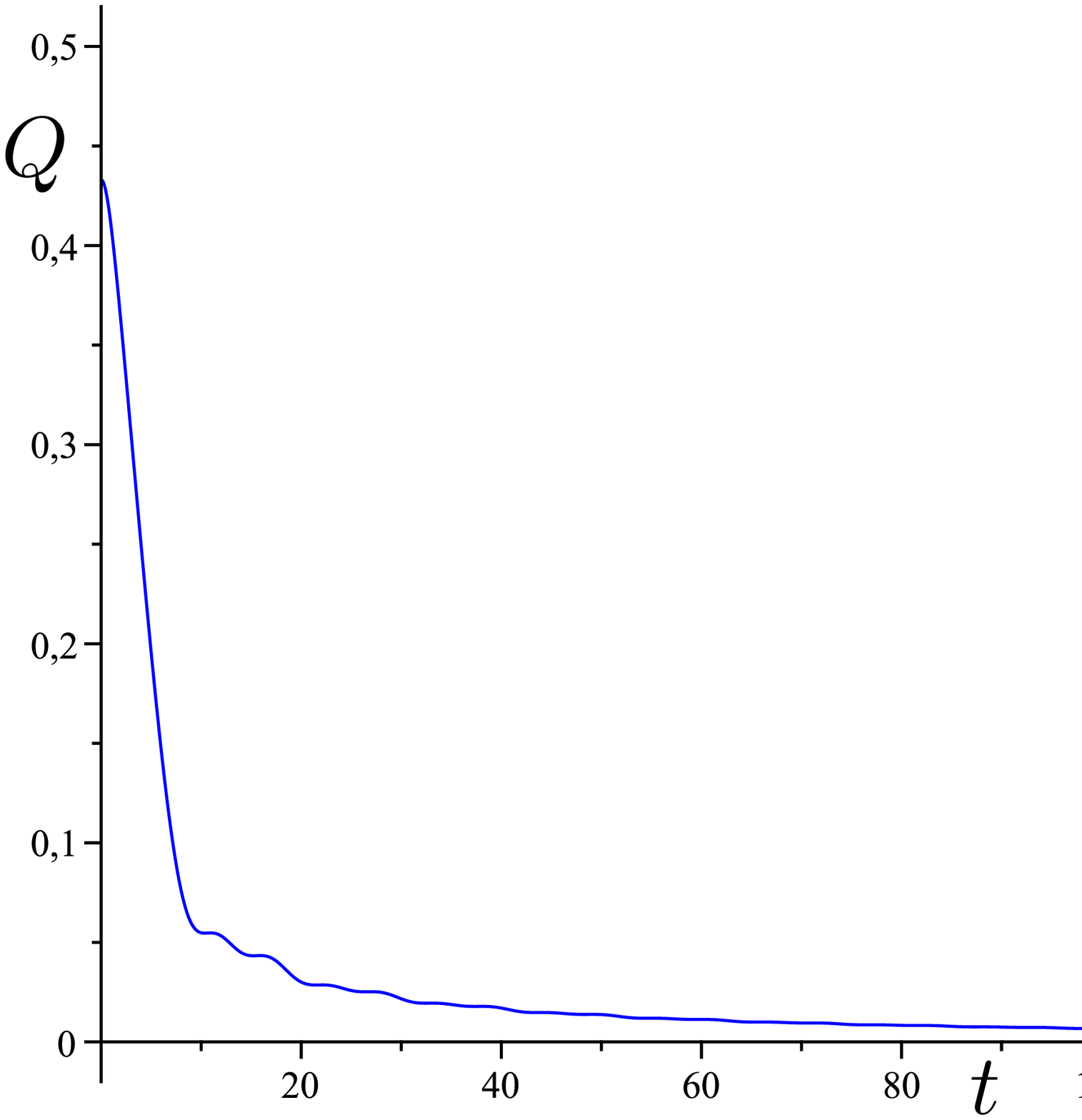}
\caption{The function $Q(t)$ at $\Lambda=0.05$ (left) and $\Lambda=0$ (right). We choose $b=1$, $\varepsilon=10$, $\xi=10$. The initial conditions are $\phi_0=2$, $\psi_0=0$, $H_0$ is calculated by (\ref{HV}) with sing "+".}
\label{Qt}
\end{figure}

Note that the obtained behaviour of the solutions is essentially different from the behaviour of the solutions in the case $\Lambda=0$.
At $\Lambda=0$ the unreachable domain is absent and the point ($H=0$, $\phi=b$, $\psi=0$) is an attractive fixed point  and the Hubble parameter is positive at any point (see Fig.~\ref{PhPlLambda0}). We also can see the difference in behaviour of the function $Q(t)$ (Fig.~\ref{Qt}).
This difference is a consequence of the unreachable domain existence.

To estimate the period of the cyclic stage we can estimate a period of oscillations for a limit trajectory --- a trajectory which coincides with a boundary of the unreachable domain $(1+6\xi)\psi^2={}-2V(\phi)$. We remind that this trajectory is not a solution for the system of equations under consideration. A solution to equation
\begin{equation*}
\ddot{\phi}=-\frac{1}{1+6\xi}V'(\phi),
\end{equation*}
with negative energy $E=\Lambda'-\varepsilon'b^4/4$, where $\Lambda'={\Lambda}/({1+6\xi})$, $\varepsilon'=\varepsilon/(1+6\xi)$, has the form (see, for example, \cite{APV})
\be
\phi(t)=A\dn(\tilde{\Omega} t+c,k),
\ee
 where $\dn(u,k)$ is the elliptic function of argument $u$ and modulus $k$, $A$ is the amplitude and $c$ is the phase. The frequency $\tilde{\Omega}$
 and the modulus $k$ of the elliptic
function  are obtained from parameters $\varepsilon'$, $b$ of the equation and depend on the
amplitude $A$:
 \bea
 \label{omega}
 \tilde{\Omega}^2&\equiv& A^2\frac{\varepsilon'}{2}\,,\qquad
 k^2\equiv 2\left(1-\frac{b^2}{A^2}\right), \\
 A^2&\equiv& b^2\left(1+\sqrt{1+\frac{4E}{\varepsilon'b^4}}\right).
 \eea
 Parameter $c$ is defined by the relation
 \be
 \phi(0)=A \dn(c,k).
 \ee

Now we estimate a period of function $\phi(t)$ (it will be of the same order as the period for the function $H$).
 Then we will also have an estimation for a period of $H$. Let us suppose the period has the standard form
\be
T=\frac{2\pi}{\tilde{\Omega}}
\ee
and substitute $\tilde{\Omega}$ in the form (\ref{omega}). We get
\be
T=2\pi \sqrt{2} \sqrt{\frac{1+6\xi}{2\sqrt{\varepsilon\Lambda}+b^2\varepsilon}}.
\ee

\subsection{The model with non-minimal coupling of the type\\ $U(\phi)=\frac{1}{2}\xi\phi^2+K$}

The model of the Higgs-driven inflation includes $R$, multiplied by $\frac{1}{2}\xi\phi^2+K$,
where $K$ is a nonzero constant.  This model with $K=M_{Pl}^2/(16 \pi)$, where $M_{\mathrm{Pl}}$ is  the Planck mass, and  $\xi=47000\sqrt{\varepsilon}$, is being considered as a real candidate to describe
the inflationary scenario~\cite{HI,Bezrukov:2008ut,HI3,HI4,Bezrukov2013}, which is fully consistent with the Planck constraints~\cite{PlanckInflation}.

  Let us consider the model with
\begin{equation}
\label{UH}
U(\phi)=\frac{\xi}{2}\phi^2+\frac{M_{\mathrm{Pl}}^2}{16\pi}.
\end{equation}
Substituting (\ref{UH}) into action (\ref{action}), we obtain:
\be
\label{action_Jordan}
S=\int d^4x \sqrt{-g} \left[\left(\frac{M_{\mathrm{Pl}}^2}{16\pi}+ \frac12 \xi \phi^2\right) R -\frac12 \partial_\mu\phi\partial^\mu \phi-V(\phi)\right].
\ee
Equations (\ref{Fr1})--(\ref{Fieldequ}) in this case are
\be
\label{e2K}
H^2=\frac{8\pi}{3(M_{\mathrm{Pl}}^2+8\pi\xi\phi^2)}\left(\frac12 \dot{\phi}^2-6\xi H\phi \dot{\phi}+V
\right),
\ee
\begin{equation}
\label{Equ11K}
3H^2+2\dot{H}=\frac{{}-8\pi}{M_{\mathrm{Pl}}^2+8\pi\xi\phi^2}\left[\frac12 \dot{\phi}^2+2\xi \dot{\phi}^2+2\xi\ddot{\phi}\phi+4\xi H\phi \dot{\phi}-V\right]\!,
\end{equation}
\begin{equation}
\label{Equ_phiK}
 \ddot{\phi}+3H\dot{\phi}+V'(\phi)+\frac{16\pi\xi\phi}{M_{\mathrm{Pl}}^2+8\pi\xi\phi^2}\left(\frac12 \dot{\phi}^2 +3\xi \dot{\phi}^2+3\xi\ddot{\phi}\phi+9\xi H\phi \dot{\phi}-2V\right)=0.
\end{equation}

Combining equations (\ref{e2K})--(\ref{Equ_phiK}) we get the system of the first order differential equations describing dynamics of the model:
\be
\label{2K}
\dot{\phi}=\psi\,,
\ee
\be
\dot{\psi}={}-3H\psi-\frac{(1+6\xi)\psi^2}{(A+6\xi)\phi}+\frac{4V-\phi A V'}{\phi(A+6\xi)},
\label{e3K}
\ee
\bea
\label{3K}
\dot{H}={}-\frac{A(1+2\xi)+4\xi}{2\xi\phi^2A(A+6\xi)}\psi^2+\frac{4H}{A\phi}\psi-\frac{4V-\phi A V'}{\phi^2A(A+6\xi)}, \eea
where $A\equiv\frac{M_{\mathrm{Pl}}^2}{8\pi\xi\phi^2}+1$.

Equation (\ref{e2K}) is a quadratic equation for $H$
and has the following solutions
\be\label{HV2}
H_\pm={}-\frac{1}{A}\frac{\psi}{\phi}\pm\sqrt{\frac{1}{A\xi}\left(\frac{6\xi+A}
{6A}\left(\frac{\psi}{\phi}\right)^2+\frac{V}{3\phi^2}\right)}.
\ee

As in the previous case (the induced gravity model) if the potential is non-positive definite then there is a possibility of a transition between $H_{+}$ and $H_{-}$.

\begin{figure}[!h]
\centering
\includegraphics[width=77mm]{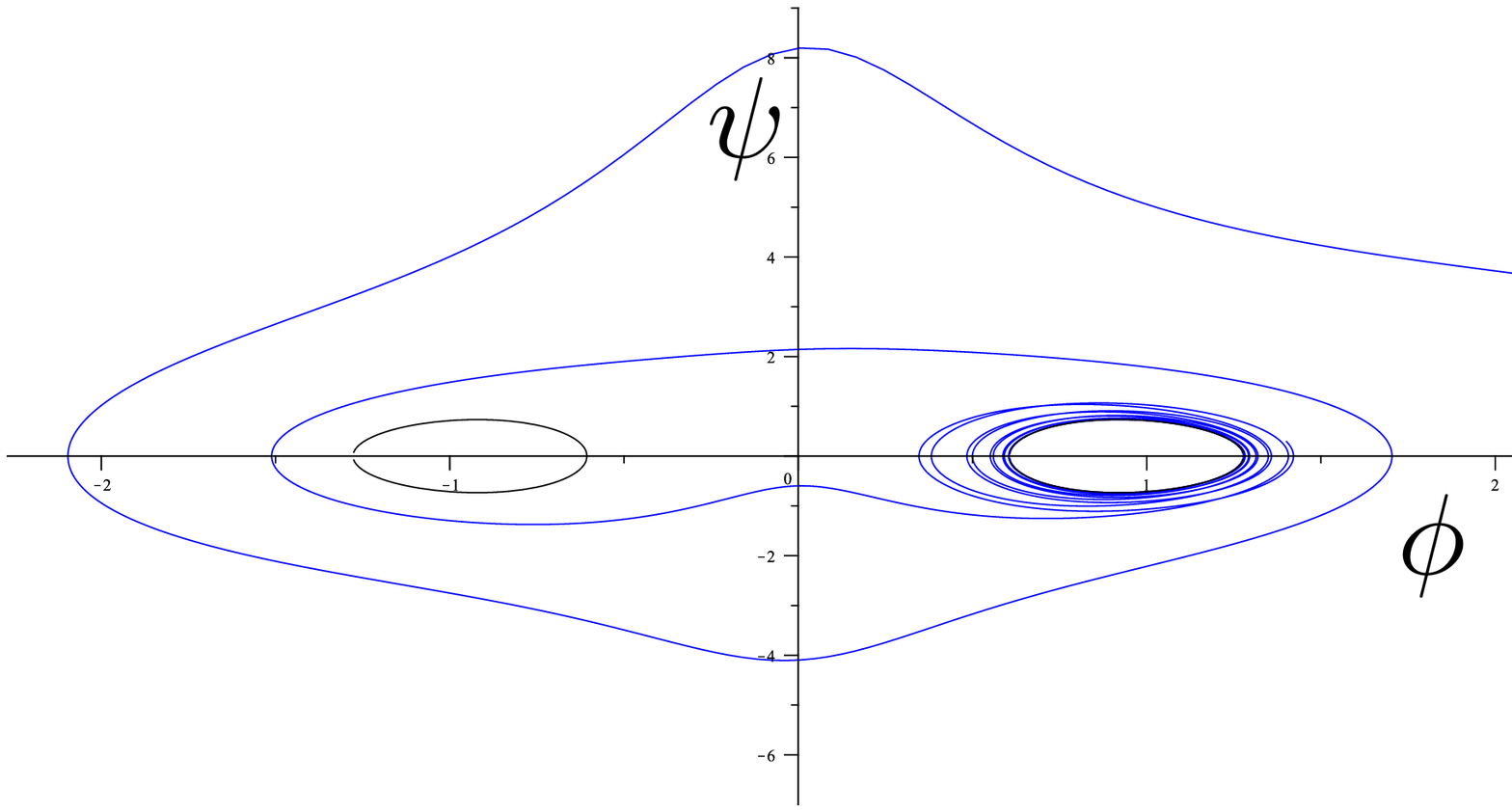}\  \  \  \  \  \  \
\includegraphics[width=45mm]{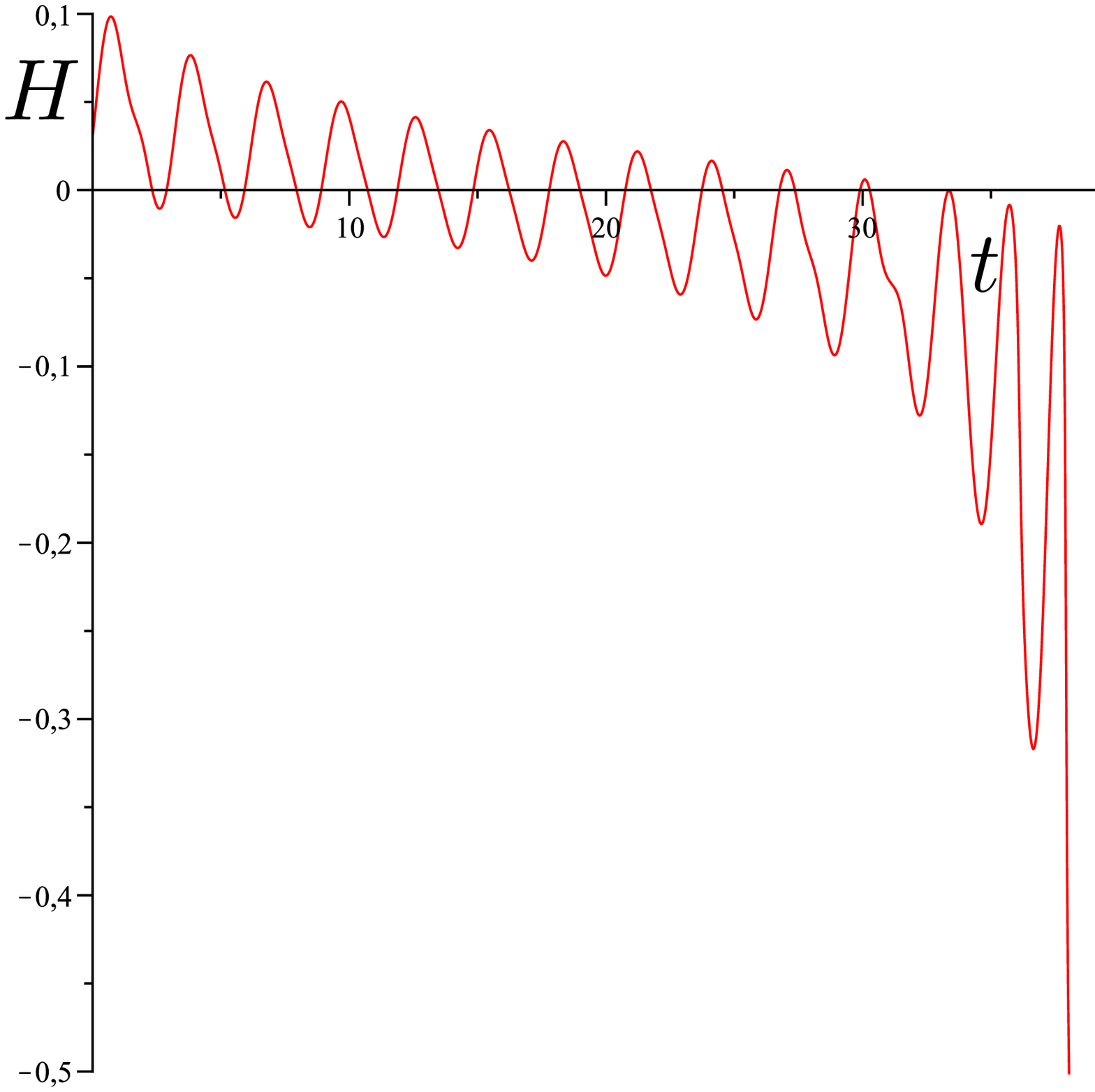}\\
\includegraphics[width=67mm]{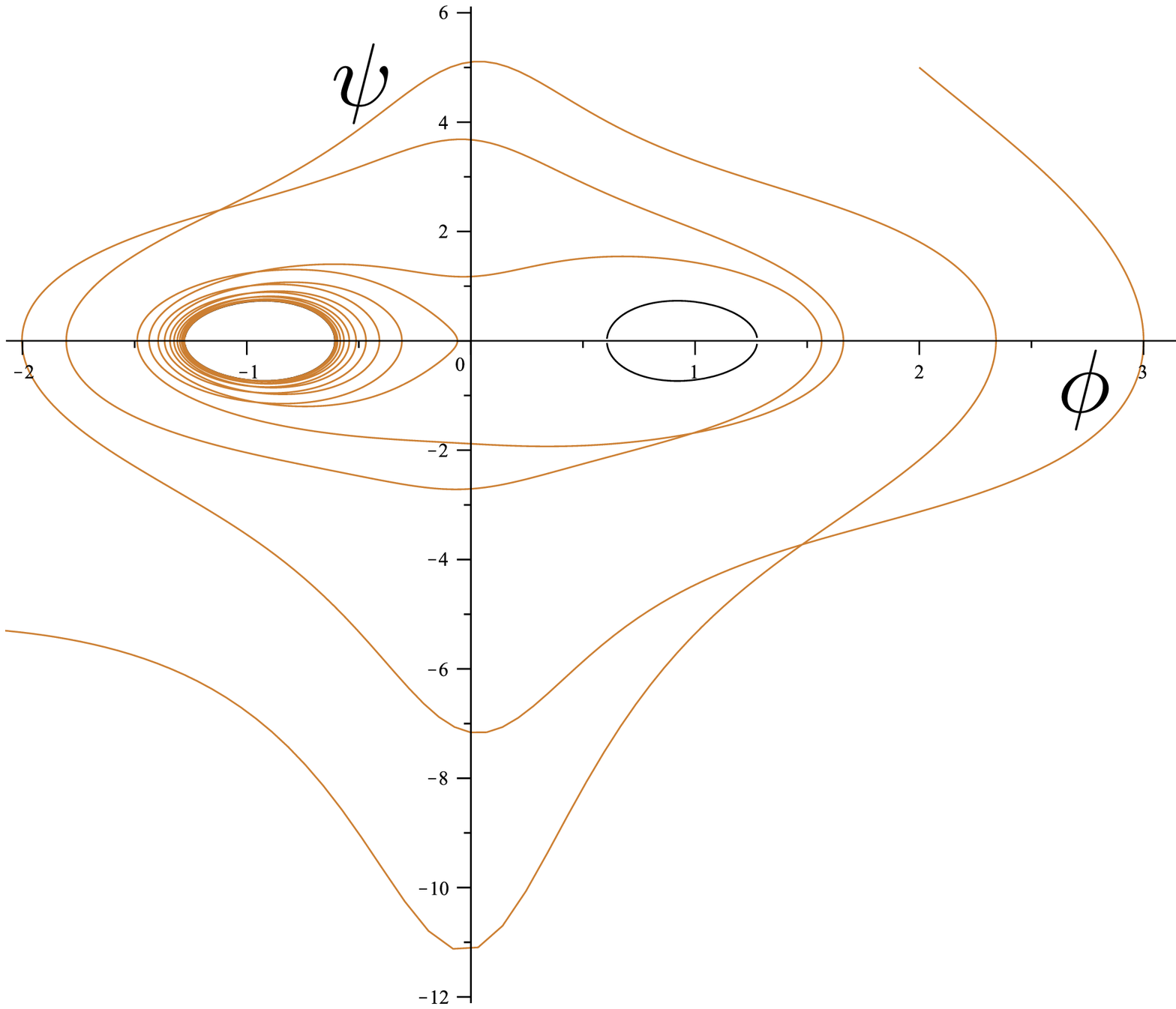}\  \  \  \  \  \  \
\includegraphics[width=55mm]{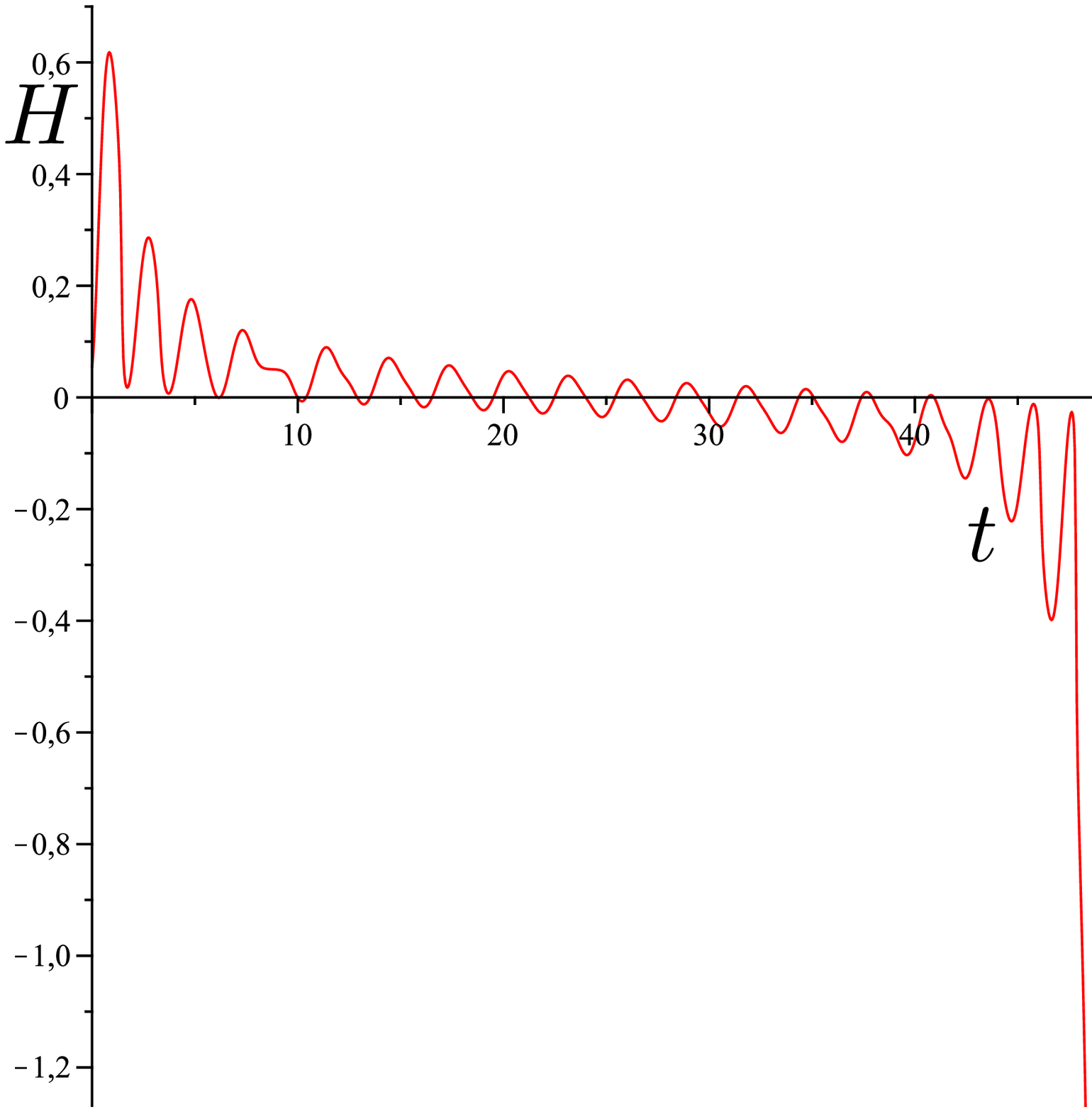}
\caption{The solution of system (\ref{2K})--(\ref{3K}) at $\Lambda=1$ and $M_{\mathrm{Pl}}=70$. We choose $b=1$, $\varepsilon=10$, $\xi=10$. The initial conditions are $\phi(0)=1.4$, $\psi(0)=0.3$ for the upper line and $\phi(0)=2$, $\psi(0)=5$ for the lower line. $H_0$ is always calculated by (\ref{HV2}) with sing "+".} \label{Pase_traj_H_i_1}
\end{figure}

We consider the non-positive definite Higgs-like potential (\ref{V_H}).
Numerical calculations give the following phase trajectory for this model (see~Fig.~\ref{Pase_traj_H_i_1}, left pictures). As in the case of the induced gravity there are two stages of evolution on the phase plane. The first stage corresponds to some kind of the cyclic Universe, then the phase trajectory reaches the boundary of the unreachable domain (inside which the values of the Hubble parameter are image) and the second stage starts. The significant difference from the induced gravity model is that on this stage of evolution the Hubble parameter doesn't decrease rapidly, but goes on oscillating evolution corresponding to the cyclic Universe. Moreover, first the phase trajectory rotates around one of the unreachable domains in a such way that the distance to the boundary of this domain increases with time, then the trajectory crosses the line $\phi=0$ and starts rotating around both unreachable domains moving farther. Then at some point the Hubble parameter rapidly decreases and the Universe contracts (see~Fig.~\ref{Pase_traj_H_i_1}, right pictures).

We can use the same approach as in the previous subsection and show that if a trajectory rotates around one of the unreachable domains starting from $H_{+}$ then it will surely reach the boundary of the unreachable domain for a finite period of time. When the trajectory touches the boundary of the unreachable domain the evolution which has been corresponding to $H_{+}$ changes to the evolution corresponding to $H_{-}$.

\section{Action and equation of motion in the Einstein frame}

Once a modified gravity theory is recast into its scalar-tensor
presentation, it immediately follows that both the Jordan frame (where
the scalar fields non-minimally couples to gravity) and the Einstein
one (which has the same form as that of Einstein gravity with minimally
coupled scalar fields) are available.
These two frames are related by conformal transformation
\begin{equation}
g_{\mu\nu}=\Omega^2 g_{\mu\nu}^{(E)}\qquad \rightarrow \qquad g=\Omega^8 g^{(E)}\,,\label{tildeg}
\end{equation}
where we denote the metric in the Jordan frame by ${g}_{\mu\nu}$,
while the one in the Einstein frame is labeled as
${g}^{(E)}_{\mu\nu}$. In the following, we denote all quantities
in the Einstein frame by adding a tag $~^{(E)}$ or  the index ${}_E$ to the corresponding
ones in the other frame, in order to avoid confusion.

Let us  consider the conformal transformation (\ref{tildeg}).
 Using $g^{\mu\nu}=\Omega^{-2}
{g^{\mu\nu}}^{(E)}$, one obtains the relationship between the Ricci
scalars in the two frames~\cite{Book-Capozziello-Faraoni,CL}
\begin{equation}
\label{conformtrans} {R}=
\Omega^{-2}\left[R^{(E)}-6\left(\Box^{(E)}\ln{\Omega}
+g^{\mu\nu(E)}\nabla^{(E)}_\mu\ln{\Omega}\nabla^{(E)}_\nu\ln{\Omega}\right)
\right].
\end{equation}
Inserting this into action (\ref{action_Jordan}),
one identifies the conformal factor as
\begin{equation}
 \Omega^{-2}=\frac{\kappa^2}{2}U \qquad \rightarrow \qquad \Omega=\frac{\sqrt{2}}{\kappa\sqrt{U}}, \label{Omega}
\end{equation}
where ${\kappa}^2 \equiv 8\pi/{M_{\mathrm{Pl}}}^2$.

The issue, which of the conformal frames, Jordan or Einstein, is the physical one, has been the
subject of longstanding debate (see~\cite{frames}, and references therein).
At the same time, it is known~\cite{Chiba:2008ia} that the spectrum of curvature perturbations and the amplitude of gravitational waves obtained are invariant under conformal transformations. By this reason it is possible to consider either frame as a physical one to study inflation and inflationary calculations are often performed in the Einstein frame. In this paper, we consider the conformal transformation from the mathematical point of view to connect the parameter $Q$ with the Hubble parameter in the Einstein frame and clarify the sense of inequality (\ref{dQ_phi}).

As is well known~\cite{Book-Capozziello-Faraoni}, conformally flat
metrics are mapped into each other. The FLRW metric is conformally
flat, so starting from the FLRW metric in the Jordan frame we obtain
the corresponding FLRW metric in the Einstein one. This leads us to
directly start from an FLRW metric with cosmic time in the Einstein
frame
\begin{equation}
\label{powersol}
ds^2={}-dt^{2}_{E}+a^{2}_{E}(t_{E})\delta_{ij}dx^idx^j\,,
\end{equation}
where in $dt_{E}$ and $a_{E}(t_{E})$, the index ${}_E$ denotes the
corresponding quantities in the Einstein frame. We get
\begin{equation}
\label{a_dt_E} dt_E=
\Omega^{-1}dt=\frac{\kappa\sqrt{U}}{\sqrt{2}} dt, \qquad a_E=\frac{\kappa\sqrt{U}}{\sqrt{2}} a\,,
\end{equation}
\begin{equation}
\label{HE}
H_E\equiv\frac{d\log{a_E}}{dt_E}=\Omega\left(H-\frac{\dot\Omega}{\Omega}\right)=
\frac{\sqrt{2}}{\kappa\sqrt{U}}\left(H+\frac{\dot U}{2U}\right)
=\frac{\sqrt{2}}{\kappa\sqrt{U}}Q.
\end{equation}

We see that the function $Q$ is connected with the Hubble parameter in the Einstein frame.
We get from (\ref{HE}) that $Q=0$ is equivalent to $H_E=0$.
As known, for cosmological models with a minimally coupling scalar field the Hubble parameter
$H_E$ is a monotonically decreasing function, therefore,
the trajectory cannot touch the boundary of unreachable domain twice. Also, it gives us the physical sense of inequality (\ref{Fr21Qm}):
\begin{equation}
\frac{dH_E}{dt_E}=\Omega\frac{dH_E}{dt}=\frac{2}{\kappa^2\sqrt{U}}\frac{d}{dt}
\left[\frac{Q}{\sqrt{U}}\right]={}-\frac{U+3{U'}^2}{2\kappa^2U^3}\dot\phi^2<0,
\end{equation}
at $U(\phi)>0$. Note that the behaviour of the Hubble parameters $H$ and $H_E$ are essentially different in the model being considered. Formula (\ref{HE}) can be useful for numeric computations of the Hubble parameter $H_E$.

\section{The special case $\xi={}-1/6$}
\label{xi_16}
This case is of a special interest due to the fact that if $\xi={}-1/6$ then equations (\ref{2})--(\ref{3}) have no sense. To consider this case we put $\xi=-1/{6}$ into equations (\ref{e2})--(\ref{Equ_phi}). These equations get the following form
\be
\label{5}
H^2+\left(\frac{\dot{\phi}}{\phi}\right)^2+2H\frac{\dot{\phi}}{\phi}={}-\frac{2V(\phi)}{\phi^2},
\ee
\be
\label{7}
3H^2+2\dot{H}={}\left(\frac{\dot{\phi}}{\phi}\right)^2-2 \frac{\ddot{\phi}}{\phi}-4H\frac{\dot{\phi}}{\phi}-\frac{6}{\phi^2}V(\phi),
\ee
\be
\label{8}
\ddot{\phi}+3H\dot{\phi}+V'+\phi\left(2H^2+\dot{H}\right)=0.
\ee

From equations (\ref{5}) and (\ref{7}) we get
\be
\label{9}
\dot{H}={}2\left(\frac{\dot{\phi}}{\phi}\right)^2 +H\frac{\dot{\phi}}{\phi}-\frac{\ddot{\phi}}{\phi}.
\ee

Using~(\ref{9}), we eliminate $\dot H$ from equation~(\ref{8}) and obtain
\begin{equation}
V'(\phi)+2\phi\left[H^2+\left(\frac{\dot{\phi}}{\phi}\right)^2+2H\frac{\dot{\phi}}{\phi}\right]=V'(\phi)-4\frac{V(\phi)}{\phi}=0.
\end{equation}

This means that a solution with a nontrivial $\phi(t)$ exists if and only if the potential $V(\phi)=V_0\phi^4$.
The constant $V_0$ must be non-positive to satisfy the requirement that $H$ is real.
Equation~(\ref{9}) is an identity in this case. We get that for any function $\phi$ there is function $H$ and they are solutions of equations.

\section{Conclusion}
We have considered the cosmological models with a non-minimally coupled scalar field and a non-positive definite potential, which is the Higgs-like potential plus a negative cosmological constant. This negative constant may appear as a result of the renormalization procedure (if there are no  protection arguments) or a result of  effective stretch of constants in SFT inspired cosmological models.

The characteristic property of models with  non-positive definite potential is the existence of
unreachable domains in the phase plane which correspond to non-real values of the Hubble parameter.
 Boundaries of
unreachable domains are given by equation (\ref{unr-boundary}).
We have shown that behaviour of the Hubble parameter in these cosmological models  with the double-well potential  essentially depends  on the sign of the  minimum  of the potential. For example, for the standard Higgs-like potential we have attractive fixed points $\phi=\pm b$, which correspond to the minima of the potential.
If we subtract a positive constant from this potential, then these fixed points turn out into the unreachable domains in the phase plane.

We have explored the dynamics of cosmological models with such non-positive definite Higgs-like potentials on the examples of the induced gravity model and model with both the Hilbert--Einstein term and the induced gravity term. We choose such values of the cosmological constant and parameters of the potential that the boundary of the unreachable domain consists of two closed curves.
We have shown numerically that the phase trajectories are being attracted to the boundary of the unreachable domain, touch it and go to infinity. It has been proved analytically that if a trajectory is attracted to the boundary of the unreachable domain, then this trajectory touches it at finite time.

The subtracting of the cosmological constant does not essentially change the behavior of model during inflation, when the scalar field changes slowly, so, in the case of non-positive definite Higgs-like potential the results should be qualitative the same: both "chaotic" inflation and "new" inflation
can be obtained.  In this context, note that the cosmological model (\ref{action_Jordan}) with the  Higgs potential considered as a real candidate to describe the inflationary scenario~\cite{HI,Bezrukov:2008ut,HI3,HI4,GB2013,Bezrukov2013}  is fully consistent with the Planck constraints~\cite{PlanckInflation}.
The substraction of the negative constant may change crucially  the reheating regime, even makes it non-realizable.
The reheating  for  the Higgs-driven inflation has been considered in~\cite{Bezrukov:2008ut}. During the reheating stage the energy stored in the scalar field is transferred to fields of the
Standard Model and maybe other (hypothetical) fields. To take the process of the scalar field decay
into account one should modify the initial equations (\ref{2})--(\ref{3}), in particular, include the decay rate $\Gamma$~\cite{induced,Allahverdi}. Also, the quantum effects play important role
(see~\cite{Allahverdi} as a review).
In this paper we do not take into account these processes and obtain that the adding of even small negative cosmological constant changes the evolution of the Universe and the Hubble parameter tends
to minus infinity instead of zero. To get the reheating for models with a non-positive definite potential
we have no choice as to provide a mechanism that compensates the negative cosmological constant at some stage of evolution, for example, by quantum effects or by switch on a new type of matter. To study all stages of cosmological evolution it is important to consider the models in presence of additional matter.
This will be a subject of our future investigations.

We have also shown that in the case of the conformal coupling ($\xi={}-1/6$) the induced gravity model has nontrivial solutions in the spatially flat FLRW metric if and only if $V(\phi)=V_0\phi^4$.

\medskip

\noindent {\bf Acknowledgements. }   We would like to thank Dmitry Gorbunov and Alexander Kamenshchik for the helpful discussions.
The work is partially supported by the RFBR grant
11-01-00894-a. The research of I.A., N.B., and R.G. is partially supported by grant NS -- 4612.2012.1.
 I.A. and R.G. are also partially supported by the Program of RAS "Mathematical Methods of the Nonlinear
Dynamics". N.B. and R.G. are also partially supported by  the RFBR grant 12-01-31144.
The research of S.V. is supported in
part by the Russian Ministry of Education and Science under grants NSh--3920.2012.2 and NSh--3042.2014.2.

\appendix
\section{The SFT motivation of the Higgs-like potential}
Let us  remind how the stretch appears and how we get~\cite{ABG} a non-positive definite Higgs-like potential, in other words, the Higgs potential minus a positive constant. Following~\cite{Aref'eva:2004qr}, we consider the string field theory (SFT) inspired nonlocal action
\begin{equation}
\label{SFT-action} S_{nl} =\!\int\! d^4x \sqrt{-g} \left[ \frac{m_{p}^2}{2} R +
 \frac{1}{2} \phi \left(\Box+\mu^2\right) e^{-\lambda\Box }\phi
-V(\phi)-\Lambda  \right]\!   ,
\end{equation} where $V(\phi)=\varepsilon \phi^4/4$, the covariant d'Alembert operator for a scalar field
 \begin{equation*}
 \Box = \frac{1}{\sqrt{-g}} \partial_\mu
\left( \sqrt{-g} \, g^{\mu
  \nu}\partial_\nu \right),
\end{equation*}
 $\phi$ is a dimensionless
scalar field and  all constants ($m_{p}$, $\mu$, $\lambda$, $\Lambda$, and $\varepsilon$) are dimensionless. A homogeneous scalar field on a spatially
flat FLRW universe with interval (\ref{friedmangeneral}) satisfies the following nonlocal equations:
\begin{equation}
\label{nonlocequ}
e^{\lambda
(\partial_t^2+3H\partial_t)}\left(\partial_t^2+3H\partial_t
-\mu^2\right)\phi(t)=-V'_{\phi},\qquad
3H^2=\frac{1}{m_{p}^2}{\cal E}.
\end{equation}

 The energy of the nonlocal field is:
\begin{equation*}
 {\cal E}=\frac{1}{2}\sum_{n=1}^\infty
\frac{\lambda^n}{n!}\sum_{l=0}^{n-1}\partial_t\Box^l\phi\,\partial_t\Box^{n-1-l}\phi+\frac{1}{2}\sum_{n=2}^\infty
\frac{\lambda^n}{n!}\sum_{l=1}^{n-1}\Box^l\phi\,\Box^{n-l}\phi-\frac{\mu^2}{2}\phi^2+\frac{\varepsilon}{4} \phi^4+\Lambda.
\end{equation*}

In the SFT inspired nonlocal cosmological models the value of the cosmological constant is fixed by the Sen conjecture~\cite{Sen} (see also~\cite{Arefeva:2001ps}) that  puts the potential energy in the nontrivial vacua equal to zero.  In other words, we get the following  potential
\be
\tilde{V}(\phi)=\frac 14 \varepsilon\phi^4-\frac 12 \mu^{2}\phi^2
+ \Lambda=\frac 14 \varepsilon\left(\phi^2-\frac{\mu^2}{\varepsilon}\right)^2,
\ee
that gives  $\Lambda=\mu^4/(4\varepsilon)$.


It has been shown in~\cite{Aref'eva:2007uk,KV,AI-IV-NEC,AK:2007,MN:2008,Vernov} that in the case of a quadratic potential a cosmological model with one nonlocal
scalar field can be presented as a model with finite or infinite number of local scalar fields and
quadratic potentials. For the nonlocal model with a quadratic potential it has been observed~\cite{Lidsey,BarCline} that in the regime suitable for the early Universe the effect of the effective stretch of the kinetic terms
takes place (see also \cite{Aref'eva:2007uk}). One can generalize this result to the case of the quartic potential and assume that
a nonlocal model can be described by  an effective local theory
with the following energy density:
\be \label{cal-E-approx} {\cal E}\approx
e^{\lambda \omega^2}\left(\frac12 \dot \phi^2-\frac 12
\mu^{2}\phi^2\right) +\frac 14 \varepsilon\phi^4 +\frac{\mu^4}{4\varepsilon},\ee
where $\omega$ is a  modified  "frequency" of the asymptotic
expansion  for equation (\ref{nor-nl-H}) in the flat case.
In the first approximation $\omega=\mu$.

Equation~(\ref{nonlocequ}) becomes
\begin{equation}
 \label{nor-nl-H}
\ddot{\phi}+3H\,\dot{\phi} -\mu^{ 2}\phi
=-\varepsilon e^{-\lambda \omega^2}\phi^3,
\end{equation}
\begin{equation}
\label{H-eff}
H^2=\frac{1}{3 m_{p\ eff}^2}\left(\frac12
\, \dot{\phi}^2+\frac{\varepsilon _{eff}}{4}\,\phi^4-\frac12\,\mu^2\phi^2+
 \Lambda_{eff}\right),
\end{equation}
where
\begin{equation*}
  m_{p\ eff}^2=m_p^2\,e^{-\lambda \omega^2},\qquad
\varepsilon_{eff}=\varepsilon e^{-\lambda \omega^2},\qquad
\Lambda_{eff}= \Lambda e^{-\lambda \omega^2}=\frac{\mu^4}{4\varepsilon_{eff}}e^{-2\lambda \omega^2}.
\end{equation*}

As known~\cite{Arefeva:2001ps}, in the SFT inspired models $\lambda >0$, so, we get the non-positive definite Higgs-like potential:
\begin{equation}
\label{Omega-shift}
  V_{L}(\phi)=\frac{\varepsilon_{eff}}{4}\,\phi^4-\frac12\,\mu^2\phi^2+
 \Lambda_{eff}=
\frac{\varepsilon_{eff}}{4} \left(\phi^2-\frac{\mu^{2}}{\varepsilon_{eff}}\right)^2+
\frac{\mu^4}{4\varepsilon_{eff}}\left(e^{-2\lambda \omega^2}-1\right).
\end{equation}

\end{document}